\documentclass[]{interact}

\usepackage{epstopdf}
\usepackage[caption=false]{subfig}

\usepackage[doublespacing]{setspace}
\usepackage{url}
\usepackage{hyperref}
\usepackage{lmodern}
\usepackage{xcolor}
\usepackage{lineno}

\usepackage[numbers,sort&compress]{natbib}
\bibpunct[, ]{[}{]}{,}{n}{,}{,}

\theoremstyle{plain}

\theoremstyle{definition}

\theoremstyle{remark}

\begin{document}

\articletype{ARTICLE TEMPLATE}

\title{Decadal analysis of sea surface temperature patterns, climatology, and anomalies in temperate coastal waters with Landsat-8 TIRS observations}


\author{
\name{Yiqing Guo\textsuperscript{a}\textsuperscript{*}, Nagur Cherukuru\textsuperscript{b}, Eric Lehmann\textsuperscript{a}, Xiubin Qi\textsuperscript{c}, Mark Doubell\textsuperscript{d}, S.~L.~Kesav Unnithan\textsuperscript{b}, Ming Feng\textsuperscript{e}}
\affil{\textsuperscript{a}CSIRO Data61, Acton, ACT 2601, Australia; \\ \textsuperscript{b}CSIRO Environment, Acton, ACT 2601, Australia; \\ \textsuperscript{c}CSIRO Space and Astronomy, Kensington, WA 6151, Australia; \\
\textsuperscript{d}South Australian Research and Development Institute, West Beach, SA 5024, Australia; \\ \textsuperscript{e}CSIRO Environment, Crawley, WA 6009, Australia;\\
\textsuperscript{*}Corresponding author: Yiqing Guo (yiqing.guo@csiro.au) 
} 
}

\maketitle

\begin{abstract}
Sea surface temperature (SST) is a fundamental physical parameter characterising the thermal state of sea surface. Due to the intricate thermal interactions between land, sea, and atmosphere, the spatial gradients of SST in coastal waters often appear at finer spatial scales than those in open ocean waters. The Thermal Infrared Sensor (TIRS) onboard Landsat-8, with its 100-meter spatial resolution, offers a unique opportunity to uncover fine-scale coastal SST patterns that would otherwise be overlooked by coarser-resolution thermal sensors. In this study, we first analysed the spatiotemporal patterns of SST in South Australia’s temperate coastal waters from 2014 to 2023 by developing an operational approach for SST retrieval from the Landsat-8 TIRS sensor. A buoy was deployed off the coast of Port Lincoln, South Australia, to validate the quality of SST retrievals. Then the daily baseline climatology of SST with 100 m resolution was constructed, which allowed for the detection and analysis of anomalous SST events. Our results suggest the following: (1) the satellite-derived SST data aligned well with the in-situ measured SST values; (2) the semi-enclosed, shallow regions of Upper Spencer Gulf and Upper St Vincent Gulf showed higher temperatures during summer and cooler temperatures during winter than waters closer to the open ocean, resulting in a higher seasonal variation in SST; (3) the near-shore shallow areas in Spencer Gulf and St Vincent Gulf, and regions surrounding Kangaroo Island, were identified to have a higher probability of SST anomalies compared to the rest of the study area; and (4) anomalous SST events were more likely to happen during the warm months than the cool months. We hope these findings would be helpful in supporting the fishing and aquaculture industries in the coastal waters of South Australia. The SST product and baseline SST climatology derived in this study have been made available for public access (please refer to the \protect\hyperlink{Data Availability}{Data Availability} section for details).
\end{abstract}


\section{Introduction}
\label{sec:introduction}

Sea surface temperature (SST) is an important geophysical variable, because it not only characterises the thermal state of sea surface but also influences the health of marine ecosystems \citep{minnett2019half, taylor2022deep, lee2025seamless}. Quantitative retrieval of SST from spaceborne measurements has been one of the primary objectives for many Earth observation missions since the 1970's \citep{legeckis1975application, legeckis1977long, legeckis1978survey}. Considering that satellite remote sensing is a powerful tool for rapid and large-scale mapping and monitoring (\emph{e.g.}, \citep{zhao2024global, guo2019nomination, yang2023trajectories}), SST maps derived from spaceborne thermal infrared instruments provide critical observational evidence to a wide range of environmental applications, including deriving water temperature climatology and detecting anomalous temperature events (\emph{e.g.}, \cite{cane1997twentieth, vecchi2007effect, pisano2020new, mohamed2022sea}).

High-quality SST retrieval from top-of-atmosphere thermal infrared measurements requires accurate correction of atmospheric effects \cite{minnett2019half,vanhellemont2020automated}. Water-emitted thermal radiance at sea level differs from that measured by spaceborne sensors due to the interference of atmosphere \cite{minnett2019half}, which should be appropriately corrected, in order to recover the thermal signals radiated by sea water \cite{vanhellemont2020automated}. Considering that the circumstance of atmosphere shifts from time to time and from location to location, it would be preferred to utilise specific atmospheric conditions at the time and location of each satellite image acquisition, rather than those sampled from a standard atmosphere database, to ensure the quality of atmospheric correction.

Although Landsat satellites are primarily designed for observing Earth's landmasses, the image footprints are able to cover sea waters close to coastline, making them suitable for coastal water monitoring. In particular, the Thermal Infrared Sensor (TIRS) onboard NASA/USGS Landsat-8 has been deemed as adequate for quality retrieval of SST \cite{vanhellemont2020automated,fu2020split,lee2022immense,vanhellemont2022validation,xie2024estimation}. The instrument measures thermal radiance at two separate spectral channels, \emph{i.e.}, 10.60--11.19 \textmu m (Band 10) and 11.50--12.51 \textmu m (Band 11) \cite{loveland2016landsat}. Despite that the
two-channel design allows a split-window approach to be applied, previous studies have reported that single-channel algorithms \cite{jimenez2006error,jimenez2008revision} based on rigorous radiative transfer modelling demonstrated superior accuracy in surface temperature retrieval from TIRS observations over both land (\emph{e.g.}, \citep{yu2014land}) and water (\emph{e.g.}, \citep{vanhellemont2020automated}). The TIRS has a spatial resolution of 100 m \cite{loveland2016landsat}, finer than the thermal bands of earlier Landsat satellites and several other spaceborne instruments like MODIS (1 km) \cite{xiong2015terra}, VIIRS (750 m) \cite{cao2013early}, and Sentinel-3 SLSTR (1 km) \cite{luo2020validation}. Given that the spatial gradients of SST in coastal waters often present at smaller spatial scales than those in open ocean waters, high-resolution SST data such as those derived from TIRS may help enhance our understanding of coastal SST patterns.

The temperate coastal waters of South Australia, including Spencer Gulf, St Vincent Gulf, and their adjacent waters, are ecologically significant and economically vital \cite{tanner2019potential}. This region supports several local industries, including fishing and aquaculture, and is home to a range of marine species, such as the endangered Great White Shark \cite{bruce1992preliminary}. Informed decision-making for local communities and industries depends on high-quality SST climatology data specific to South Australia's coastal waters. The SSTAARS (Sea Surface Temperature Atlas of the Australian Regional Seas) dataset \cite{wijffels2018fine} has been a reliable source of baseline SST climatology (\emph{e.g.}, \cite{huang2021high, huang2024marine, pena2022revisiting, hemming2023observed}), but its spatial resolution, at a relatively coarse 2 km, may limit its applicability in studies requiring finer-scale oceanographic details. In particular, the complex meandering coastline and small-scale SST features in South Australia's coastal waters necessitate a finer spatial resolution for SST climatology to capture the localised thermal dynamics. Previous studies reported that, fine-scale thermal gradients at submesoscale resolution (0.1--10 km), together with larger thermal features such as mesoscale jets and eddies (10--100 km), form key drivers of marine ecosystem response in coastal waters \cite{gommenginger2019seastar, petrenko2017review}. The 100 m resolution TIRS observations thus present an opportunity to extract SST climatology with enhanced spatial detail that would support the analysis of ecological processes and inform aquacultural decisions. 

Anomalous SST events that deviate from the baseline climatology impact fishery and aquaculture practices, as well as the health of marine ecosystems. South Australia's coastal waters are influenced by the interaction of large-scale ocean currents, such as the warm Leeuwin Current Extension flowing from the Indian Ocean along the southern coastline \cite{cresswell1980observations} and the cooler Flinders Current originating from the Southern Ocean \cite{bye1978oceanic}. The two large triangular-shaped gulfs, namely the Spencer Gulf and St Vincent
Gulf, enclose waters characterised by regional circulation. These geographical factors, along with land-water interactions and vertical upwelling dynamics, contribute to the high frequency of SST anomalies in South Australia's coastal waters, as reported in \cite{streten1981southern,vivier2010dynamics}. To assist local communities and industries in mitigating the potential adverse effects of anomalous SST events, a clearer understanding of their spatiotemporal distribution is essential.

In this study, we focus on analysing the SST patterns, climatology, and anomalies in temperate coastal waters off the South Australian coast with high-resolution satellite observations for the ten-year period from 2014 to 2023. We develop several operational approaches for high-resolution SST monitoring in South Australian coastal waters, including:
\begin{enumerate}
\item Implementing an operational approach for high-quality SST retrieval from the Landsat-8 TIRS sensor, with rigorous radiative transfer modelling and spatiotemporally specific atmospheric data being applied for correcting atmospheric effects;
\item Establishing the daily baseline SST climatology at a spatial resolution of 100 m for South Australian coastal waters; 
\item Compiling the probability map of anomalous SST events in the study area with SST images over 2014--2023, and analyse the spatial and seasonal distribution of these anomalies; and
\item Implementing a tile-based parallelisation scheme for efficient processing across large spatial scales, with the derived SST product and baseline SST climatology for the study area being made available for public access (please refer to the \protect\hyperlink{Data Availability}{Data Availability} section for details on data access). 
\end{enumerate}

Through quantitative analysis of the data generated from the above developments, we sought to answer primarily the following questions: (1) What are the spatial and seasonal patterns of SST in South Australian coastal waters? (2) What are the characteristics of SST climatology in the study area? (3) Which regions within the study area experienced more frequent anomalous SST events during 2014--2023?

The rest of the paper is organised as follows. Section~\ref{sec:study_area_and_data_sets} introduces the study area and the datasets used for modelling and validation. Section~\ref{sec:methods} describes the methodologies for SST retrieval and spatiotemporal analysis. Section~\ref{sec:results_and_discussions} presents the validation and analysis results, and discusses the findings and limitations of this work. Finally, Section~\ref{sec:conclusion} concludes the paper. 
\section{Study Area and Data Sets}
\label{sec:study_area_and_data_sets}

\subsection{Study Area}

\begin{figure}[!htb]
\centering
\includegraphics[width=13cm]{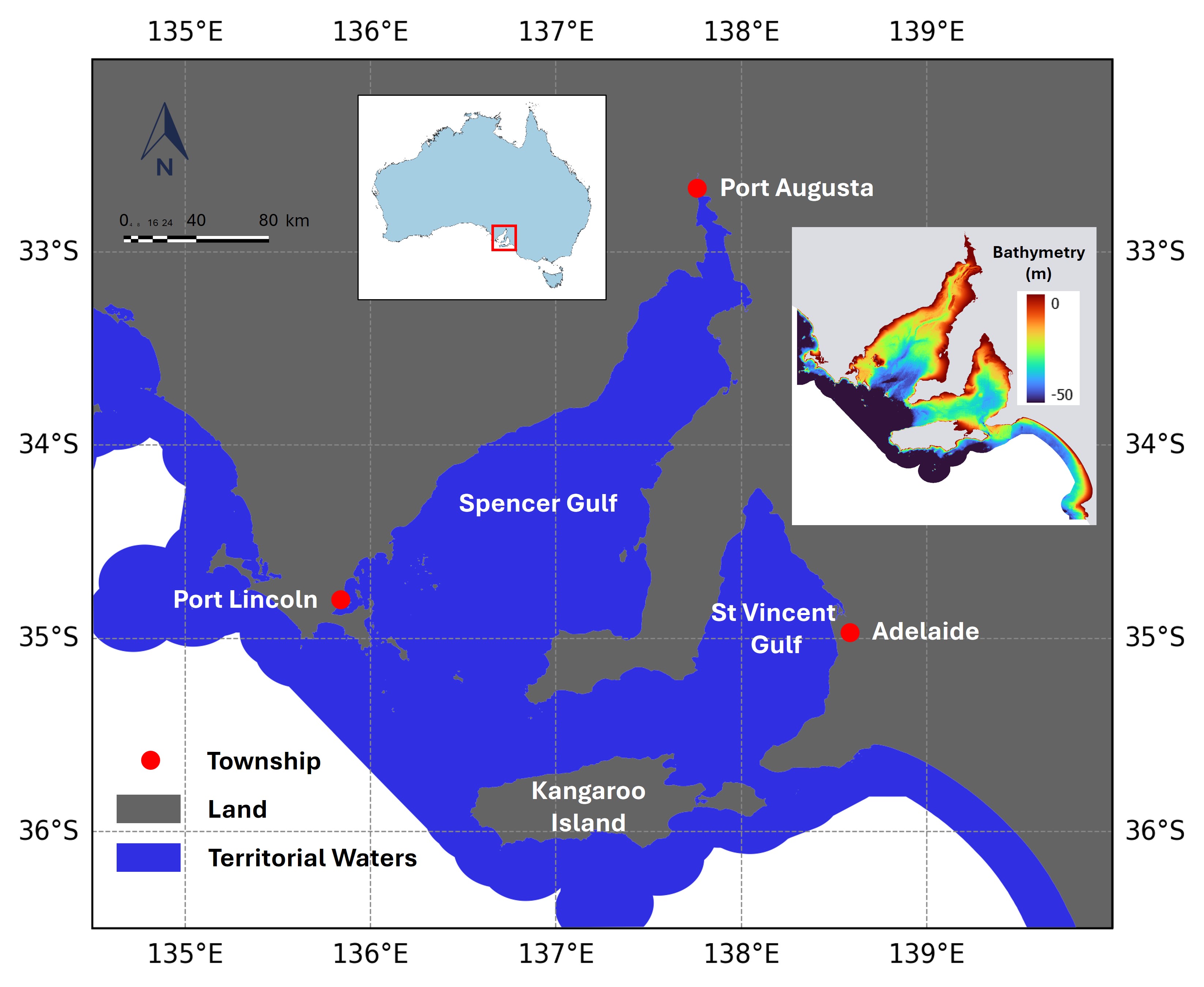}
\caption{The study area of this work encompasses the Australian territorial waters off the coast of South Australia, including Spencer Gulf, St Vincent Gulf, and their adjacent waters.\label{fig:location}}
\end{figure}

This work was focused on Australian territorial waters off the coast of South Australia (134\textdegree30'00''--140\textdegree00'00''E, 32\textdegree25'58''--36\textdegree30'00''S), including Spencer Gulf, St Vincent Gulf, and their adjacent waters, as shown in Figure~\ref{fig:location}. The South Australian territorial waters were chosen as the study area because they are ecologically significant as habitat for species such as the endangered Great White Shark, and economically vital due to their role in supporting commercial fisheries and aquaculture industries. This coastal region falls within the spatial coverage of the Landsat-8 TIRS sensor. The shapefiles defining the coastline and territorial water limit were obtained from the Australian Bureau of Statistics\footnote{\href{https://www.abs.gov.au/statistics/standards/australian-statistical-geography-standard-asgs-edition-3/jul2021-jun2026/access-and-downloads/digital-boundary-files}{https://www.abs.gov.au/statistics/standards/australian-statistical-geography-standard-asgs-edition-3/jul2021-jun2026/access-and-downloads/digital-boundary-files}}. These shapefiles contain high-resolution vector data, which were rasterised to a 100-m grid to ensure consistency with the resolution of TIRS satellite imagery. Major townships surrounding the study area include Adelaide (the capital city of South Australia, located on the east bank of St Vincent Gulf), Port Lincoln (located on the west bank of Spencer Gulf), and Port Augusta (situated at the north end of Spencer Gulf) (Figure~\ref{fig:location}). The Kangaroo Island is located to the south of Spencer Gulf and St Vincent Gulf (Figure~\ref{fig:location}). This temperate coastal region has a Mediterranean climate, with hot and dry summers, and mild and wet winters. Air temperatures range from 18.5{\textdegree C} to 29.2{\textdegree C} in January, and from 7.2{\textdegree C} to 16.9{\textdegree C} in July. The area receives moderate rainfall, averaging 277.5 mm per year, with most of it falling during the winter months, according to statistics from Australian Bureau of Meteorology\footnote{\href{http://www.bom.gov.au/climate/averages/tables/cw\_018103.shtml}{http://www.bom.gov.au/climate/averages/tables/cw\_018103.shtml}}.

\subsection{Satellite Imagery}

\begin{figure}[!htb]
\centering
\includegraphics[width=10cm]{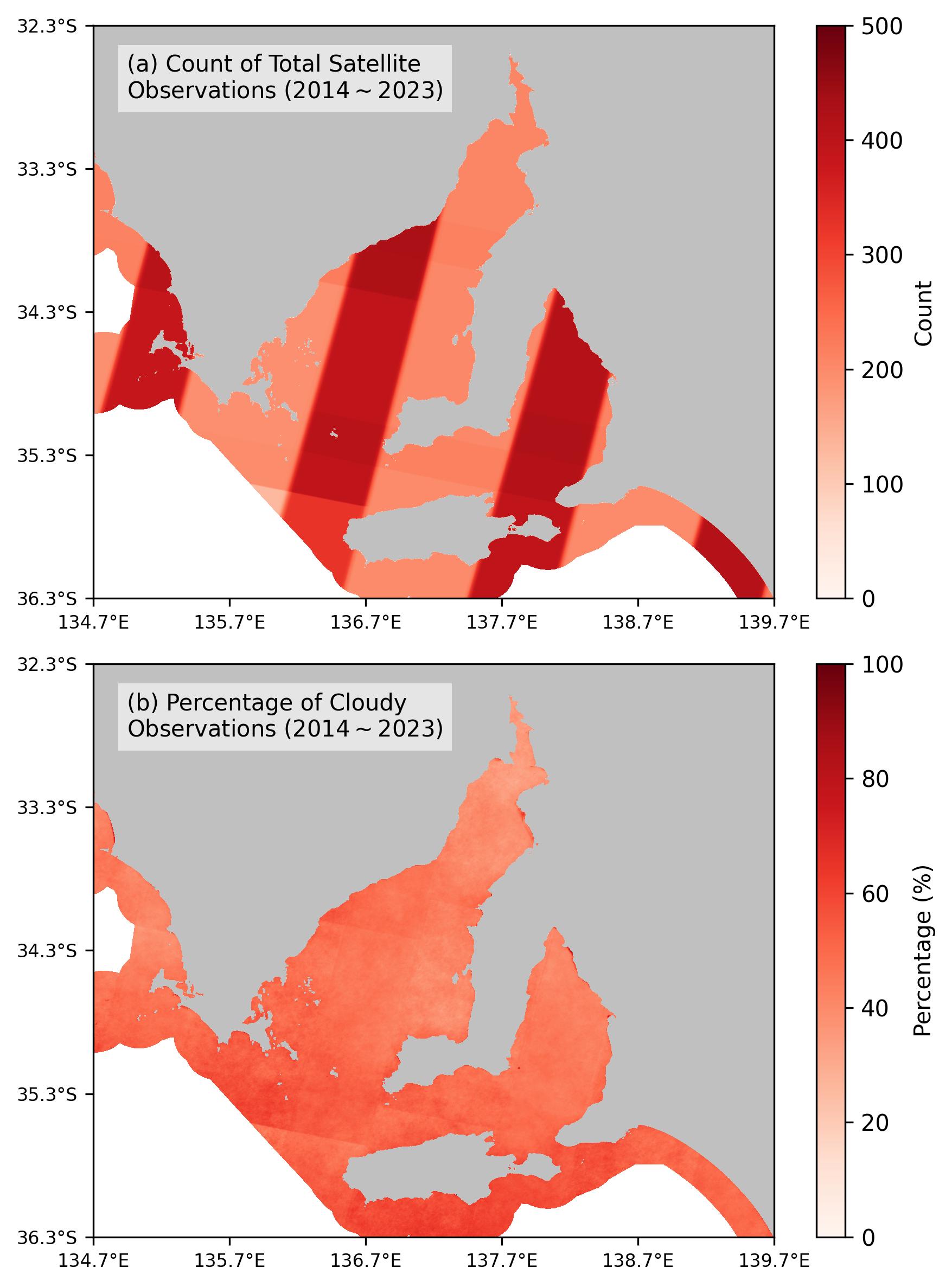}
\caption{(a) The number of total satellite observations in the study area over 2014--2023, and (b) the chance of cloudy observations over the same time period as derived from the Landsat Collection 2 Pixel Quality Assessment Band. \label{fig:count}}
\end{figure}

As the thermal infrared instrument onboard NASA/USGS Landsat-8 satellite, TIRS is capable of measuring thermal radiation emitted from Earth's surface. It records thermal radiance at two separate spectral channels, \emph{i.e.}, 10.60--11.19 \textmu m (Band 10) and 11.50--12.51 \textmu m (Band 11), with a spatial resolution of 100 m. The TIRS has a swath width of 185 kilometres, aligning with the coverage of Operational Land Imager (OLI) onboard the same satellite. In this study, the Landsat-8 Collection 2 Level-1C TIRS product was used. It consists of top-of-atmosphere thermal infrared data with radiometrical calibration and geometrical correction having been applied. TIRS images covering the study area from 2014 to 2023 were queried within the AquaWatch Data Service (ADS) built upon CSIRO’s Earth Analytics and Science Innovation (EASI) platform, where an Open Data Cube (ODC) environment \cite{lewis2017australian,killough2018overview} was set up to facilitate data processing. The study period of 2014--2023 was selected because 2014 marks the first full year of operational-quality data from the TIRS sensor onboard Landsat-8 following its launch in 2013, while 2023 represents the most recent year for which the EAC4 dataset was available at the time of this study. The total number of satellite observations from 2014 to 2023 varies across the study area, ranging from 127 to 431 records, as shown in Figure~\ref{fig:count}a. Regions with overlapping satellite scenes show a higher observation frequency. Cloud cover is frequent in the study region, as shown in Figure~\ref{fig:count}b, with the percentage of cloudy observations ranging from 30.6\% to 72.9\%. The lowest cloud percentage is observed in the Upper Spencer Gulf, and the highest over the ocean south of Kangaroo Island. In this study, pixels flagged as cloud or cirrus by the Landsat Collection 2 Pixel Quality Assessment Band were masked out from our processing. In addition to these Landsat-8 TIRS images, we also queried Landsat-8 OLI Level-2 surface reflectance images and MODIS Terra Level-2 SST images for quality control and intercomparison purposes. 

\subsection{Coincident Atmosphere Data}

\begin{figure}[!htb]
\centering
\includegraphics[width=14.3cm]{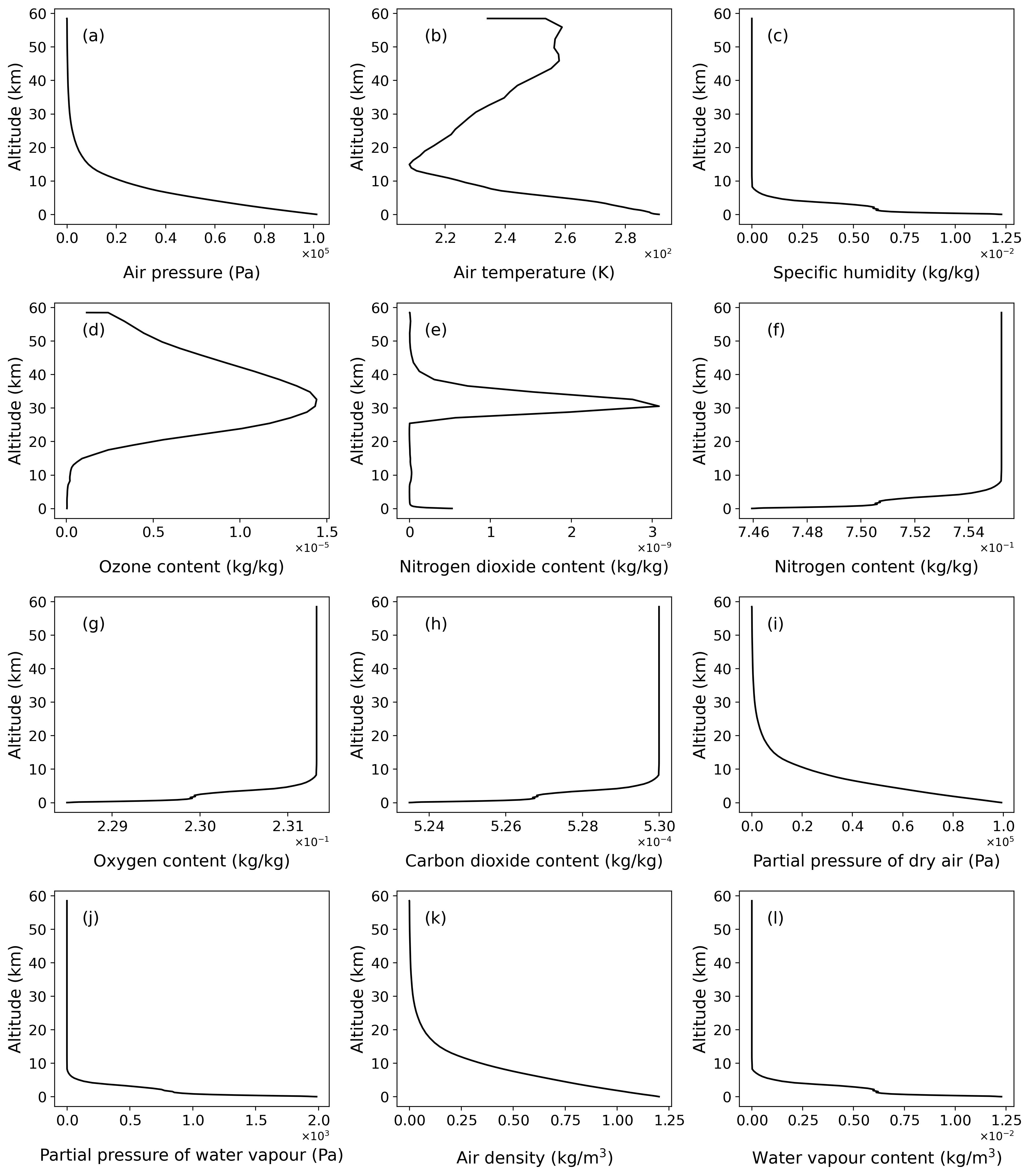}
\caption{Example vertical profiles of (a) air pressure, (b) air temperature, (c) specific humidity, (d) ozone content, (e) nitrogen dioxide content, (f) nitrogen content, (g) oxygen content, (h) carbon dioxide content, (i) partial pressure of dry air, (j) partial pressure of water vapour, (k) air density, and (l) water vapour content from the sea surface (0 km altitude) to the top of atmosphere (60 km altitude). \label{fig:atmos}}
\end{figure}

The atmosphere data spatiotemporally coincident with each satellite observation were retrieved from the CAMS global reanalysis (EAC4) dataset \cite{inness2019cams}, provided by the Copernicus Atmosphere Monitoring Service (CAMS) at the European Centre for Medium-Range Weather Forecasts (ECMWF). We chose to use spatiotemporally specific atmosphere data at the time and location of each satellite observation, rather than the data of a standard atmosphere, aiming to improve the modelling accuracy of SST retrievals. The EAC4 dataset provides high-quality atmosphere data at a 3-hour temporal interval and a spatial resolution of approximately 80 km. These data were queried using the Climate Data Store Application Program Interface (CDSAPI) from the Atmosphere Data Store, where the data were provided in an interpolated 0.75{\textdegree}{\texttimes}0.75{\textdegree} lon/lat grid. The vertical profiles of air pressure (in unit of Pa; Figure~\ref{fig:atmos}a), air temperature (in K; Figure~\ref{fig:atmos}b), specific humidity (in kg/kg; Figure~\ref{fig:atmos}c), ozone content (in kg/kg; Figure~\ref{fig:atmos}d), and nitrogen dioxide content (in kg/kg; Figure~\ref{fig:atmos}e) were retrieved. These profiles were provided at 60 vertical levels of the standard atmosphere model as defined by the United States Standard Atmosphere, 1976 \cite{atmospheric1976us}. It is worth noting that the model levels are associated with air pressure rather than elevation. We therefore interpreted these model levels according to \cite{documentation2020part}, and converted them into altitudes (in km relative to the sea level). The vertical profiles of non-trace gasses, including nitrogen, oxygen, and carbon dioxide contents (all in kg/kg; Figures~\ref{fig:atmos}f--h), were calculated based on their percentage constitutions in dry air moderated by the vertical distribution of specific humidity. The vertical profiles of water vapour pressure and dry air pressure (both in Pa; Figures~\ref{fig:atmos}i--j) were derived from air pressure and specific humidity based on the physical equations documented in \cite{murray1967computation, shaman2009absolute}. The vertical profile of air density (in kg/m\textsuperscript{3}; Figure~\ref{fig:atmos}k) was then derived from dry air pressure and water vapour pressure based on the Ideal Gas Law. Finally, the vertical profile of water vapour content (in kg/m\textsuperscript{3}; Figure~\ref{fig:atmos}l) was converted from air density and specific humidity. These atmosphere data were applied to SST modelling in this study.

\subsection{Ground Truth Measurements}

\begin{figure}[!htb]
\centering
\includegraphics[width=13cm]{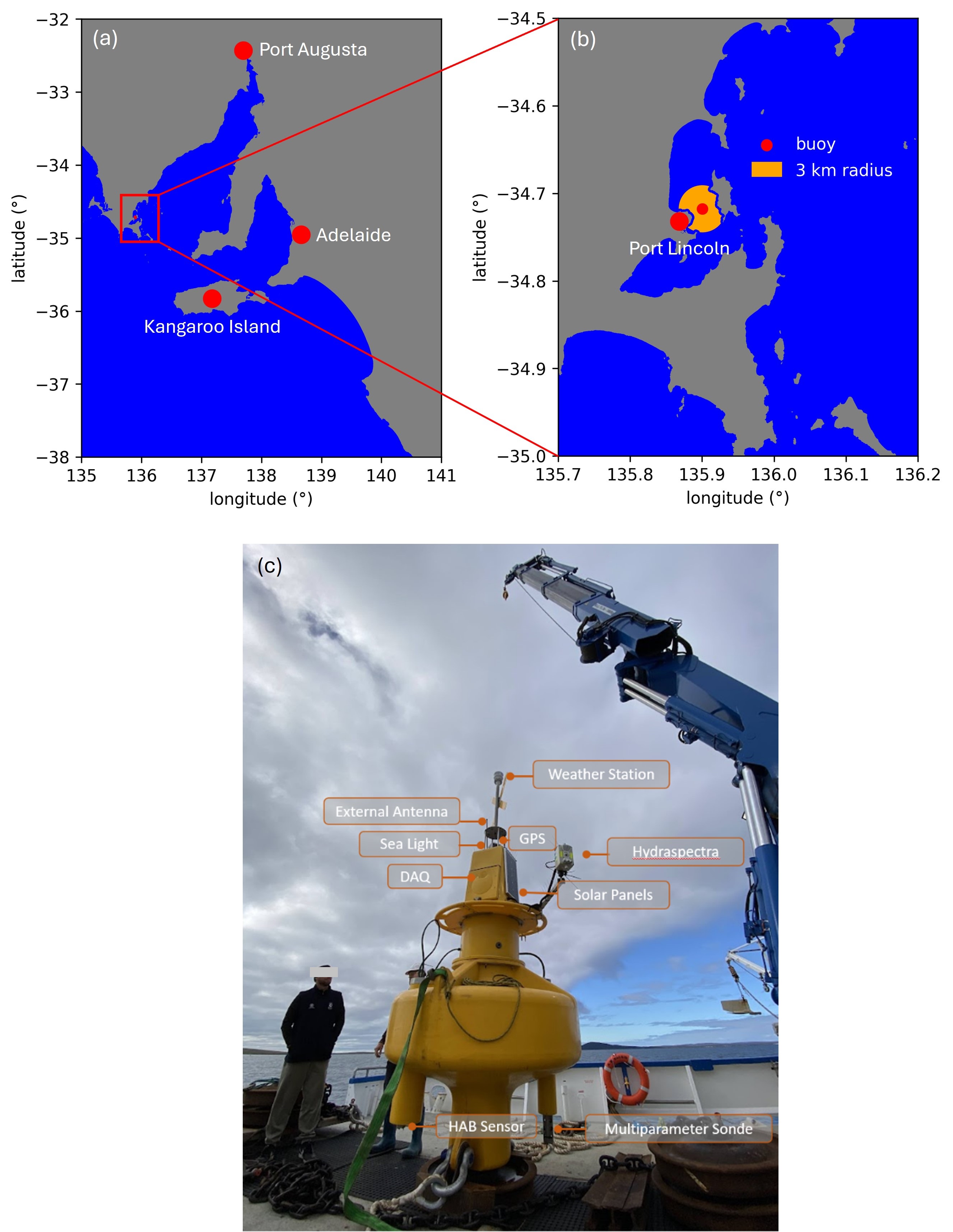}
\caption{A buoy has been set up near Port Lincoln in this study to collect in-situ sea surface temperature (SST) for validating the satellite-derived SST data. (a) Location of the buoy; (b) Zoomed-in view of the buoy location; and (c) A photo of the buoy before deployment.\label{fig:buoy}}
\end{figure}

To validate the satellite-derived SST data in this study, we deployed a buoy off the coast of Port Lincoln, South Australia (135\textdegree54'01''E, 34\textdegree43'03''S), with support from CSIRO's AquaWatch Mission, as shown in Figure~\ref{fig:buoy}. A YSI EXO2 multiparameter sonde is mounted onto the buoy which streams in-situ bulk temperature every 10 min at a nominal depth of 1 m. The in-situ sensor has been in operation since 31 August 2022, with regular services and calibrations being conducted to ensure the quality of the data. These in-situ temperature measurements served as a reference for validating the satellite-derived SST data. For each satellite observation, the derived SST pixels covering a buffer zone of 3 km radius (excluding land pixels and near-shore pixels within 300 m to the coastline; Figure~\ref{fig:buoy}b), centred at the buoy location, were averaged and compared to the in-situ temperature reading closest to the acquisition time of satellite imagery. This averaging process aimed to increase the quantity of data points and provide a more robust SST estimate.

As an extended experiment to evaluate the performance of the proposed SST retrieval algorithm for temperate coastal waters beyond the study area, we sourced in-situ SST measurements from three buoys off the southeast coast of Australia \cite{OEH_Data}. These buoys, managed by the Manly Hydraulics Laboratory, provide high-quality SST records in New South Wales coastal waters. The records of bulk temperature were collected using a thermistor located at the base of the buoy approximately 0.5 meters below the water surface \cite{OEH_Data}. Table \ref{tab:wave_buoys} details the station name, station code, location, and record period for each buoy.

\begin{table}
\tbl{List of offshore wave data buoys with their locations and record periods.}
{
\begin{tabular}{llll} \toprule
Station Name                            & Station Code & Location                                                                    & Record Period \\ \midrule
Batemans Bay Offshore Buoy  & WAVEBAB      & 150°20'13"E, 35°42'28"S & 01 Apr 2013 -- 31 Dec 2015 \\
Coffs Harbour Offshore Buoy & WAVECOH      & 153°15'32"E, 30°22'22"S & 01 Apr 2013 -- 31 Dec 2017 \\
Crowdy Head Offshore Buoy   & WAVECRH      & 152°51'08"E, 31°49'26"S & 01 Apr 2013 -- 31 Dec 2017 \\ \bottomrule
\end{tabular}
}
\label{tab:wave_buoys}
\end{table} 
\section{Methods}
\label{sec:methods}

\subsection{Overview}

In this study, we aim at analysing the spatiotemporal dynamics of SST in temperate coastal waters in South Australia with satellite-derived SST measurements. We first implement an operational approach to retrieving SST from Landsat-8 TIRS observations, followed by conducting statistical analyses to examine the spatiotemporal patterns, climatology, and anomalies of SST in the study area. In Subsection~\ref{ssec:radiative_transfer_modelling}, we briefly review the well-established equations underlying the radiative transfer process from the water-emitted radiance to the sensor-captured top-of-atmosphere radiance. In Subsection~\ref{ssec:derivation}, we describe the derivation of atmospheric radiative variables via three radiative transfer simulations with the \emph{libRadtran} software package \cite{emde2016libradtran}, and the calculation of atmospheric parameters from CAMS EAC4 atmosphere data which are required as inputs for the \emph{libRadtran} simulations. Then, in Subsection~\ref{ssec:calculation_of_sst}, we calculate SST based on the radiative transfer equations and derived atmospheric radiative variables. In Subsection~\ref{ssec:parallelisation}, we introduce a framework to parallelise the SST calculation for large spatiotemporal scales. Finally, we detail the methodologies for analysing the spatiotemporal patterns of SST, deriving the baseline SST climatology, and detecting anomalous SST events in Subsection~\ref{ssec:spatiotemporal_analyses}.

\subsection{Radiative Transfer Modelling}
\label{ssec:radiative_transfer_modelling}

\begin{figure}[!htb]
\centering
\includegraphics[width=14cm]{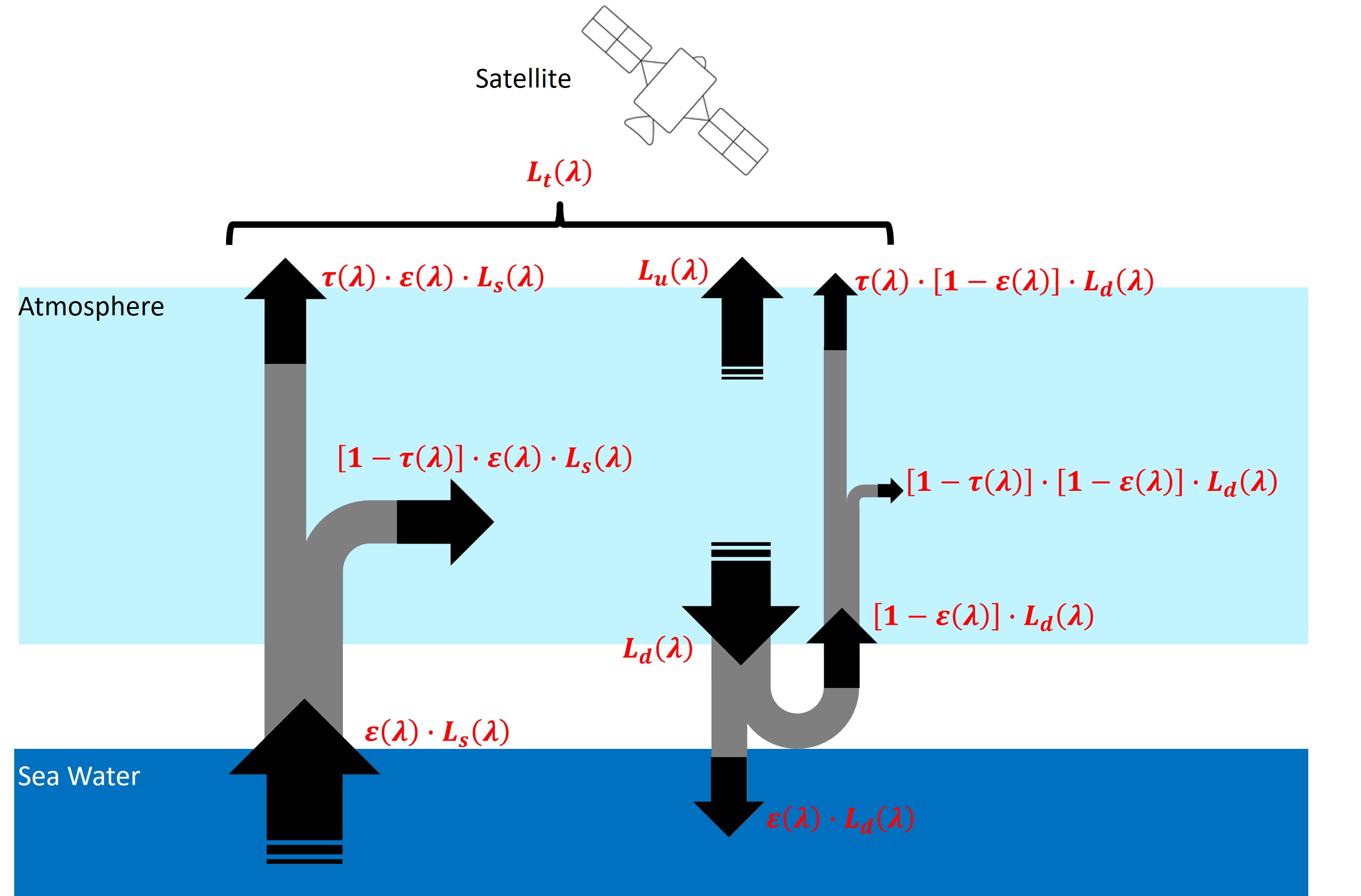}
\caption{Simplified radiative transfer process from the radiance of sea water to at-sensor radiance, with interference from atmospheric scattering, absorption, and emission. $L_s(\lambda)$ is the black-body radiance of sea water; $L_t(\lambda)$ is the at-sensor radiance; $L_u(\lambda)$ and $L_d(\lambda)$ are atmospheric upwelling and downwelling radiances, respectively; $\tau(\lambda)$ is the transmittance of atmosphere; $\varepsilon(\lambda)$ is the emissivity of sea water; and $\lambda$ denotes spectral dependency.\label{fig:radiative}}
\end{figure}

Figure~\ref{fig:radiative} illustrates the radiative transfer process from sea surface radiance to at-sensor radiance, with interference from atmospheric scattering, absorption, and emission. As depicted in the figure, the emission of thermal infrared radiance from the surface of sea water, $\varepsilon(\lambda)\cdot L_s(\lambda)$, is a product of the black-body radiance of water $L_s(\lambda)$ (which is determined by sea surface temperature $T$ through the Planck's law) and water emissivity $\varepsilon(\lambda)$, where $\lambda$ denotes spectral dependency. The water-emitted radiance then passes through the atmosphere with a transmittance coefficient of $\tau(\lambda)$. Part of the radiation, $\left[1-\tau(\lambda)\right]\cdot\varepsilon(\lambda)\cdot L_s(\lambda)$, is absorbed and scattered by atmospheric constituents such as water vapour and carbon dioxide. The remaining part, $\tau(\lambda)\cdot\varepsilon(\lambda)\cdot L_s(\lambda)$, reaches the top of atmosphere.

In addition to the water-emitted radiance, the atmosphere itself also emits radiation (Figure~\ref{fig:radiative}). The atmospheric upwelling radiance, $L_u(\lambda)$, is the radiant energy that is emitted or scattered upwards from the atmosphere. The atmospheric downwelling radiance, $L_d(\lambda)$, is that going downward, which is then partially absorbed by sea water ($\varepsilon(\lambda)\cdot L_d(\lambda)$) and the rest is re-emitted upward ($\left[1-\varepsilon(\lambda)\right]\cdot L_d(\lambda)$). The re-emitted radiation then passes upward through the atmosphere, with part being absorbed and scattered ($\left[1-\tau(\lambda)\right]\cdot\left[1-\varepsilon(\lambda)\right]\cdot L_d(\lambda)$) and the rest reaching the top of atmosphere ($\tau(\lambda)\cdot\left[1-\varepsilon(\lambda)\right]\cdot L_d(\lambda)$).

The total at-sensor radiance measurable by the thermal infrared instrument onboard the satellite, $L_t(\lambda)$, is the combination of water-emitted radiance and atmospheric upwelling and downwelling radiances that reach the top of atmosphere. The entire radiative transfer process can be described mathematically as the following:
\begin{equation}
\label{eq:radiative}
L_t(\lambda) = \tau(\lambda)\cdot\varepsilon(\lambda)\cdot L_s(\lambda) + L_u(\lambda) + \tau(\lambda)\cdot\left[1-\varepsilon(\lambda)\right]\cdot L_d(\lambda),
\end{equation}
where $L_s(\lambda)$ is the black-body radiance of sea water; $L_u(\lambda)$ and $L_d(\lambda)$ are atmospheric upwelling and downwelling radiances, respectively; $\tau(\lambda)$ is the transmittance of atmosphere; and $\varepsilon(\lambda)$ is the emissivity of sea water. The radiative transfer model in Eq.~(\ref{eq:radiative}) establishes the forward relationship between the black-body radiance of sea water, $L_s(\lambda)$, and the at-sensor radiance, $L_t(\lambda)$.

\subsection{Derivation of Atmospheric Radiative Variables}
\label{ssec:derivation}
In Eq.~(\ref{eq:radiative}), the three variables $\tau(\lambda)$, $L_u(\lambda)$, and $L_d(\lambda)$ characterise the radiative properties of atmosphere. Following the approach in \cite{vanhellemont2020automated}, three atmospheric simulations were conducted with the radiative transfer software package \emph{libRadtran} \cite{emde2016libradtran}, to derive these variables. Details of the three \emph{libRadtran} simulations are provided in Appendix~\ref{sec:appendix_a}.

The parameters of atmospheric conditions (\emph{e.g.}, air pressure, temperature, and relative humidity) were required as inputs for the three \emph{libRadtran} simulations. Instead of applying a standard mid-latitude summer or winter atmosphere as the default atmospheric conditions, we chose to make use of spatiotemporally specific atmospheric conditions when and where the satellite observation was captured. Considering that accurate atmospheric corrections are critical in SST retrieval, location- and time-specific atmospheric conditions may help improve the quality of the retrieved SST values.

In this study, we used the CAMS EAC4 atmosphere data to derive spatiotemporally specific atmospheric conditions for each satellite observation. Considering that the spatial resolution of the CAMS EAC4 dataset is 0.75 degree longitude by 0.75 degree latitude and the temporal resolution is 3 hours, we divided the TIRS image archive of interest into a spatiotemporal grid with each cell in the grid covering images over an area of 0.75 degree longitude by 0.75 degree latitude and a time period of 3 hours (as detailed later in Subsection~\ref{ssec:parallelisation}). While this method improves the alignment between the coarse-resolution EAC4 atmospheric data and the fine-resolution TIRS data, some degree of mismatch remains for areas with complex atmospheric dynamics, suggesting a need for higher-resolution atmospheric datasets in future studies. For each cell that covers a spatial span of 0.75 degree longitude by 0.75 degree latitude and a temporal span of 3 hours, the spatiotemporally specific atmosphere variables, including air pressure $p^{(i)}$, air temperature $T^{(i)}$, air density $\rho^{(i)}$, water vapour content $q^{(i)}$, nitrogen concentration $N_2^{(i)}$, oxygen concentration $O_2^{(i)}$, and carbon dioxide concentration $CO_2^{(i)}$, ozone concentration $O_3^{(i)}$, and nitrogen dioxide concentration $NO_2^{(i)}$, at 60 model levels from sea surface to the top of atmosphere, were provided as inputs to the \emph{libRadtran} simulations. Detailed calculations of these atmosphere variables are provided in Appendix~\ref{sec:appendix_b}.

\subsection{Calculation of SST}
\label{ssec:calculation_of_sst}

After determining the atmospheric radiative variables, \emph{i.e.}, $\tau(\lambda)$, $L_u(\lambda)$, and $L_d(\lambda)$, was described in Subsection~\ref{ssec:derivation}, we convolved these spectral quantities with the TIRS spectral response functions to convert them into broadband quantities that match with TIRS Band 10 (10.60--11.19 \textmu m) and Band 11 (11.50--12.51 \textmu m). We denote these broadband atmospheric radiative variables as $\tau^{\textnormal{B10}}$, $L_u^{\textnormal{B10}}$, and $L_d^{\textnormal{B10}}$ for Band 10, and $\tau^{\textnormal{B11}}$, $L_u^{\textnormal{B11}}$, and $L_d^{\textnormal{B11}}$ for Band 11. By convolving a lab measurement of the spectral emissivity of sea water with TIRS spectral response functions, the broadband emissivity of sea water at TIRS Bands 10 and 11 were determined to be $\varepsilon^{\textnormal{B10}}=0.9926$ and $\varepsilon^{\textnormal{B11}}=0.9877$, as documented in \cite{vanhellemont2020automated}. Then, deriving from Eq.~(\ref{eq:radiative}), the black-body radiance of sea water at Band 10, $L_s^{\textnormal{B10}}$, can be calculated as:
\begin{equation}
\label{eq:black_body_radiance_b10}
L_s^{\textnormal{B10}} = \frac{L_t^{\textnormal{B10}} - L_u^{\textnormal{B10}}}{\tau^{\textnormal{B10}}\cdot\varepsilon^{\textnormal{B10}}} - \frac{\left(1-\varepsilon^{\textnormal{B10}}\right)\cdot L_d^{\textnormal{B10}}}{\varepsilon^{\textnormal{B10}}},
\end{equation}
where $L_s^{\textnormal{B10}}$ is the at-sensor radiance recorded by TIRS Band 10. The same calculation can be applied to get the black-body radiance of sea water at Band 11, $L_s^{\textnormal{B11}}$.

The SST can be retrieved from either $L_s^{\textnormal{B10}}$ or $L_s^{\textnormal{B11}}$ based on the physical relationship between black-body radiance and temperature, as described by Planck's Law. In this study, SST was calculated as the average over the retrievals from both bands:
\begin{equation}
\label{eq:calculate_t}
T = \frac{1}{2} \left[ \frac{K_2^{\textnormal{B10}}}{\ln \left( \frac{K_1^{\textnormal{B10}}}{L_s^{\textnormal{B10}}} + 1 \right)} + \frac{K_2^{\textnormal{B11}}}{\ln \left( \frac{K_1^{\textnormal{B11}}}{L_s^{\textnormal{B11}}} + 1 \right)} \right] - 273.15,
\end{equation}
where $T$ is the derived SST in degree Celsius (\textdegree C); $K_1^{\textnormal{B10}}=774.8853$, $K_2^{\textnormal{B10}}=1321.0789$, $K_1^{\textnormal{B11}}=480.8883$, and $K_2^{\textnormal{B11}}=1201.1442$ are sensor-specific constants as documented in Landsat 8 Data Users Handbook \cite{usgs_landsat8_handbook}. It is worth noting that the SST values derived here represent the skin temperature of the water surface. To validate them against the bulk temperatures measured by in-situ sensors, the skin temperatures were compared with in-situ bulk temperature measurements over two years for the Port Lincoln buoy, and over 2.5 years for the New South Wales buoys. An offset of 0.6{\textdegree}C was applied to the Port Lincoln buoy and 0.9{\textdegree}C to the New South Wales buoys, to account for the difference between skin and bulk temperatures.

\subsection{Parallelisation for Large-Scale Processing}
\label{ssec:parallelisation}

\begin{figure}[!htb]
\centering
\includegraphics[width=14cm]{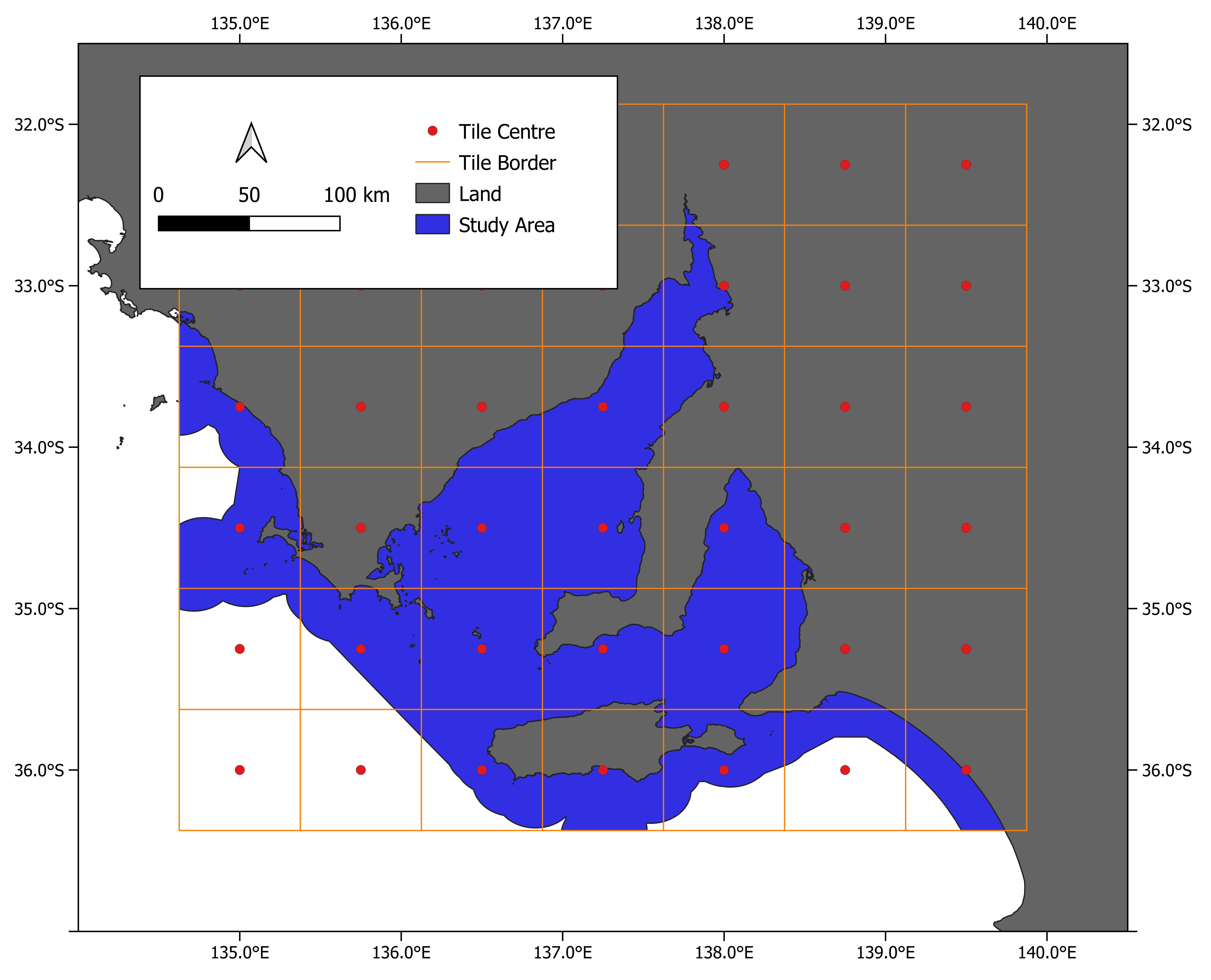}
\caption{The study area is divided into 0.75{\textdegree}{\texttimes}0.75{\textdegree} lon/lat tiles for accelerated and efficient processing, where the tile centres align with the sampling points of CAMS EAC4 atmosphere data. \label{fig:grid}}
\end{figure}

Given the spatiotemporal scale of this study, we divided the study area into multiple tiles to facilitate parallelised and efficient processing. Each tile encompasses a 0.75° × 0.75° longitude/latitude area, with its centre aligned to the sampling point of the CAMS EAC4 atmospheric data, as illustrated in Figure~\ref{fig:grid}. Multitemporal TIRS images from 2014 to 2023 were extracted and processed independently for each tile. For efficiency, only the tiles intersecting the study area were processed, while water tiles outside the study area and pure-land tiles were skipped. These tiles were distributed across the workers of a Dask cluster on Amazon Web Services within CSIRO's EASI platform to accelerate data processing via parallelised computing.

\subsection{Spatiotemporal Analyses}
\label{ssec:spatiotemporal_analyses}

To analyse the spatiotemporal patterns of SST in the study area, we computed the mean SST and coefficient of variation (C.V.) of SST from 2014 to 2023. These metrics allowed us to identify spatial patterns in SST levels across the study area as well as areas with high temporal variability. Monthly SST maps were also generated to capture seasonal temperature fluctuations. This analysis aimed to provide seasonal information on SST for supporting the local fishery and aquaculture industries. 

\begin{figure}[!htb]
\centering
\includegraphics[width=12cm]{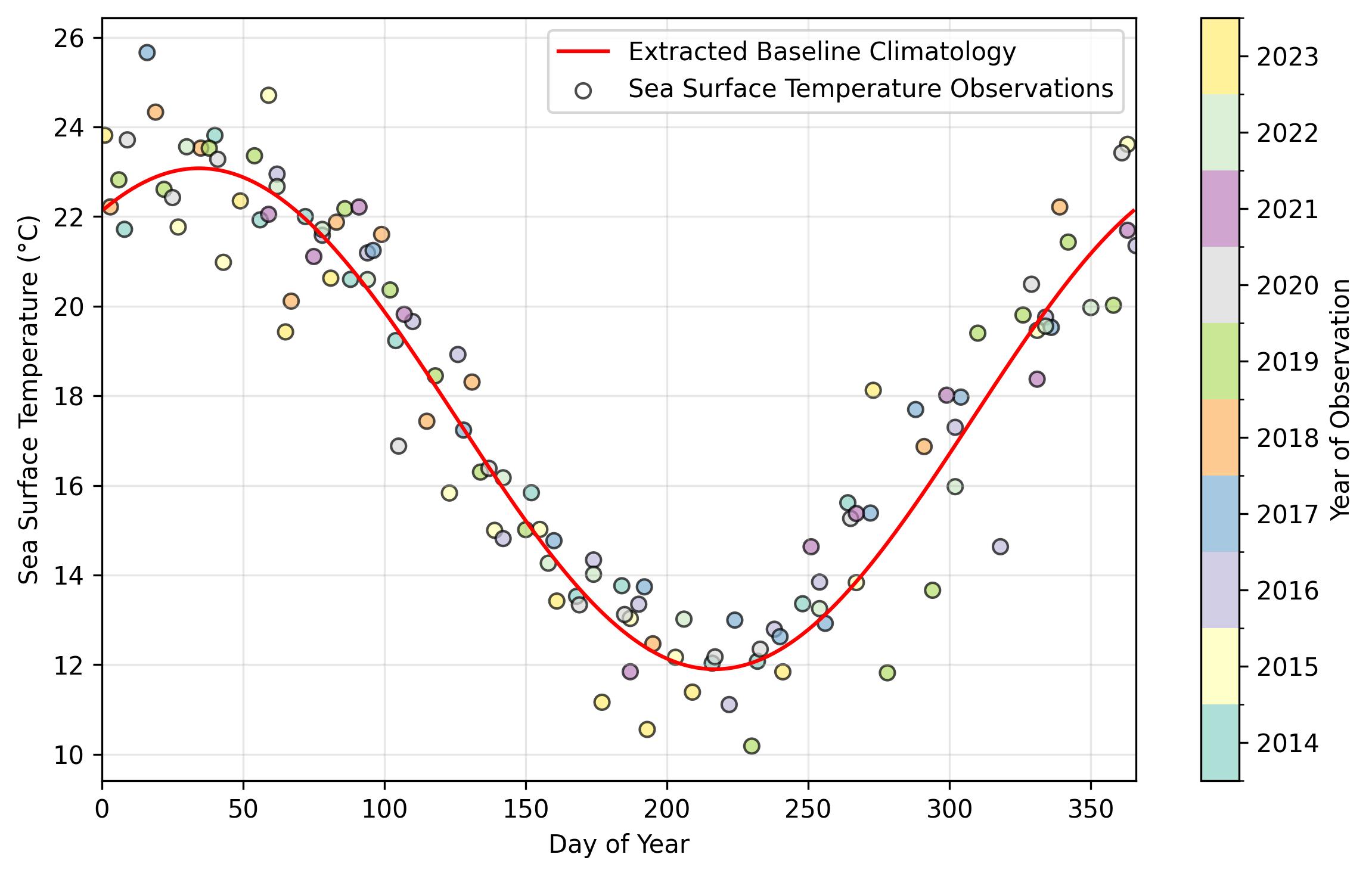}
\caption{Demonstration of baseline climatology extraction for sea surface temperature (SST) at a randomly selected location (137.6{\textdegree}E, 33.4{\textdegree}S) within the study area.\label{fig:fitting}}
\end{figure}

The baseline climatology of SST was extracted for each calendar date with the ten years of SST imagery (2014--2023) derived in this study. Specifically, for each pixel, we first amalgamated SST observations from all years into one single calendar year based on their acquisition day of the year, followed by fitting these observations to a sinusoidal function expressed as:
\begin{equation}
\label{eq:sinusoidal}
T(d) = A \cdot \cos\left( \frac{2 \pi d}{365} + \phi \right) + O,
\end{equation}
where $T(d)$ is the SST observation at the $d$th calendar day of a year. The selection of parameters, amplitude $A$, phase $\phi$, and offset $O$, was determined in this study through non-linear least-squares regression aiming at optimising the agreement between the fitted climatology and observed SST values. Figure~\ref{fig:fitting} illustrates the procedure of extracting the baseline climatology for a randomly selected pixel. This approach was applied to every pixel within the study area to generate the climatology maps.

With the baseline climatology, we calculated the probability map of anomalous temperatures in the study area, which is defined as the percentage of SST observations from 2014 to 2023 that are greater than or less than 2°C from the baseline temperature. This threshold was empirically determined based on observations reported in the literature, which indicate that a temperature offset exceeding 2{\textdegree}C may cause adverse impacts on marine species, such as increasing the risk of mass coral mortality \cite{decarlo2017mass}. However, anomaly maps can be generated under other threshold values following the same approach described here. An anomaly map was generated for the study area by calculating the frequency of temperature anomalies, aggregating both positive and negative deviations from the climatological norm. Based on the anomaly map, regions with high probabilities of anomalous SST were identified. To examine the seasonal distribution of SST anomalies, we further categorised the detected anomalous events by the warm months (January--March and October--December) and cool months (April–-September), and compared the likelihood of SST events between these two seasons.

\section{Results and Discussions}
\label{sec:results_and_discussions}

\subsection{Validation of SST Retrievals}
\label{ssec:validation_of_sst_retrievals}

\begin{figure}[!htb]
\centering
\includegraphics[width=14cm]{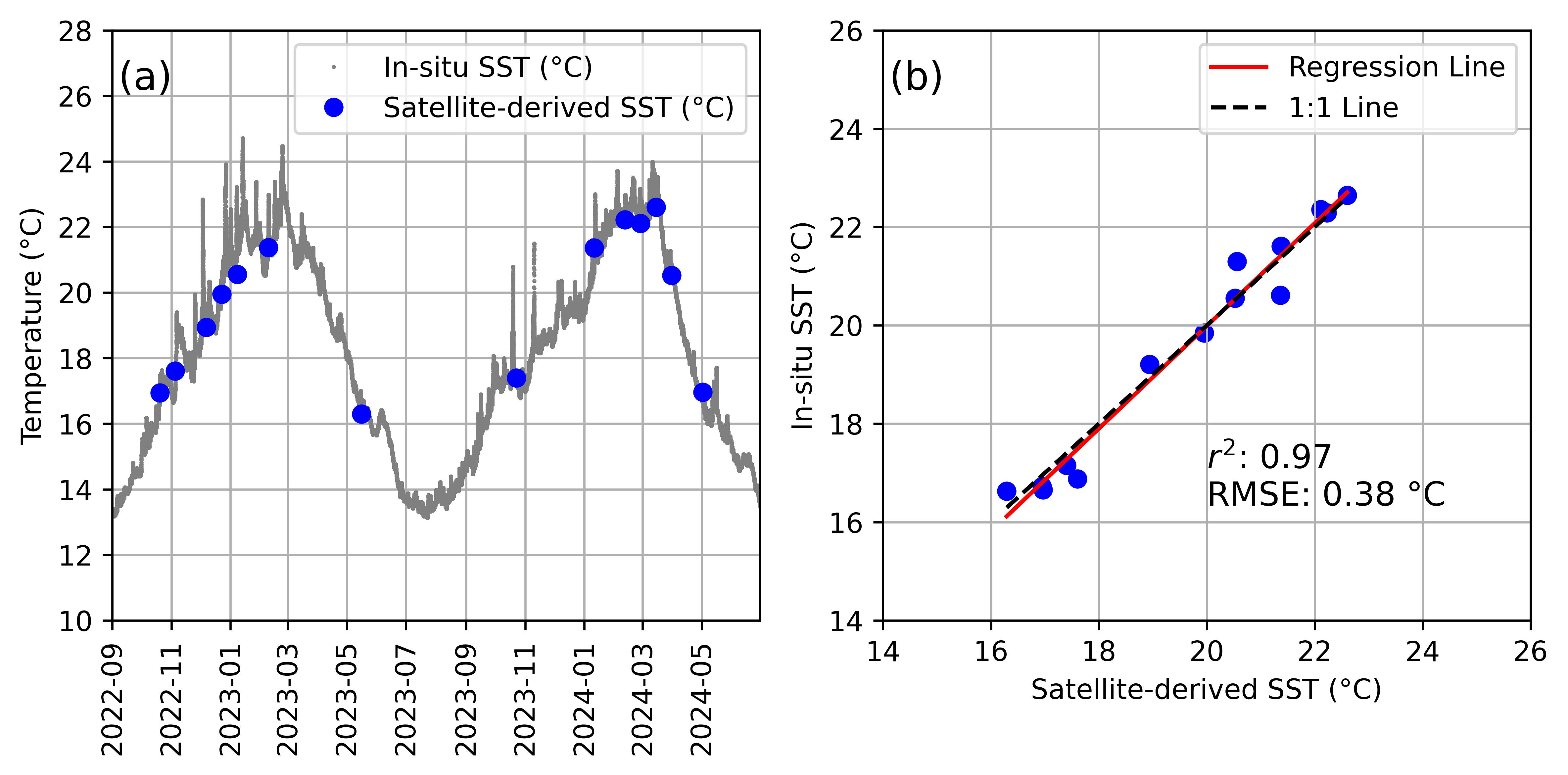}
\caption{(a) Time-series comparison and (b) scatter plot between satellite-derived sea surface temperature (SST) and in-situ SST measurements collected at the location of the Port Lincoln buoy (135\textdegree54'01''E, 34\textdegree43'03''S). \label{fig:port_lincoln}}
\end{figure}
\begin{figure}[!htb]
\centering
\includegraphics[width=13.5cm]{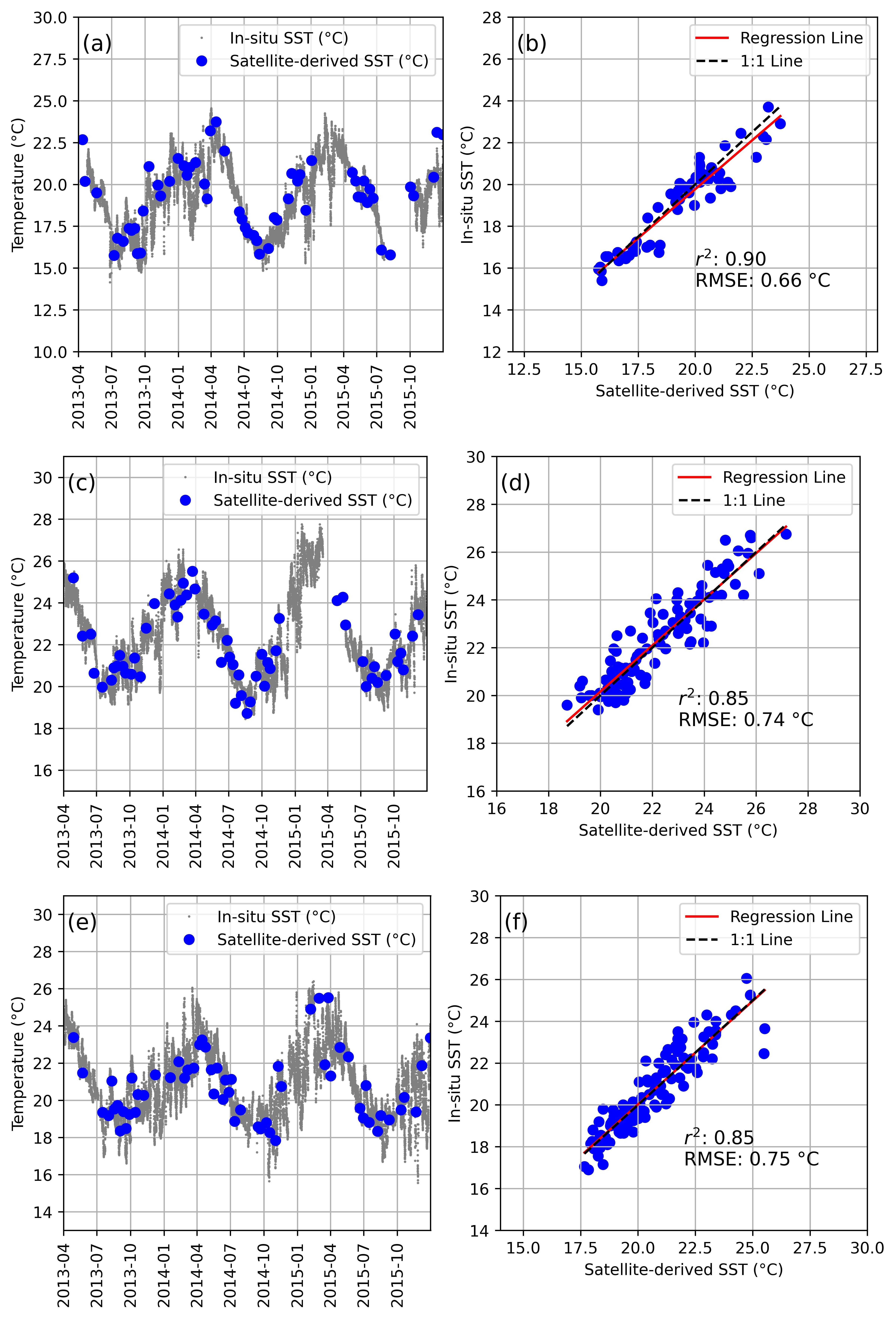}
\caption{Time-series comparisons and scatter plots between satellite-derived sea surface temperature (SST) and in-situ SST measurements at (a--b) the Batemans Bay Offshore Buoy (150°20'13"E, 35°42'28"S), (c--d) Coffs Harbour Offshore Buoy (153°15'32"E, 30°22'22"S), and (e--f) Crowdy Head Offshore Buoy (152°51'08"E, 31°49'26"S). \label{fig:validation_wave}}
\end{figure}

Figures~\ref{fig:port_lincoln} and \ref{fig:validation_wave} present the validation results of satellite-derived SST data against in-situ SST measurements. A total of 14, 57, 113, and 115 satellite/in-situ matchups were identified for the Port Lincoln Buoy, Batemans Bay Offshore Buoy, Coffs Harbour Offshore Buoy, and Crowdy Head Offshore Buoy, respectively. An in-situ SST measurement was considered as a valid match with a satellite-derived SST if the two were recorded within 100 m spatial distance and less than 30 minutes apart in time.

The time-series plot in Figure~\ref{fig:port_lincoln}a shows the comparison between satellite-derived SST data and in-situ SST measurements collected at the location of Port Lincoln buoy (135\textdegree54'01''E, 34\textdegree43'03''S). It was observed that the satellite-derived SST values aligned well with the in-situ measurements. It was also seen that the overall temperature trends, including seasonal variations and peaks, were captured by the satellite-derived SST. The scatter plot in Figure~\ref{fig:port_lincoln}b presents the relationship between the satellite-derived and in-situ SST. The quantitative assessment shows a reasonably strong correlation between the two, with an $r^2$ value of 0.97 and a root mean square error (RMSE) of 0.38{\textdegree}C. 

The results for the Batemans Bay Offshore Buoy (150°20'13"E, 35°42'28"S), Coffs Harbour Offshore Buoy (153°15'32"E, 30°22'22"S), and Crowdy Head Offshore Buoy (152°51'08"E, 31°49'26"S) are shown in Figures~\ref{fig:validation_wave}a--b, \ref{fig:validation_wave}c--d, and \ref{fig:validation_wave}e--f, respectively. It was observed that the $r^2$ values are within the range of 0.85--0.90 and the RMSE values are lower than 0.75{\textdegree}C. The validation results in Figures~\ref{fig:port_lincoln} and \ref{fig:validation_wave} suggested that the SST retrievals obtained using the algorithm proposed in this study provide a reliable estimate of the temperature of sea surface waters. 

\subsection{Visualisation of SST Images}

\begin{figure}[!htb]
\centering
\includegraphics[width=11.5cm]{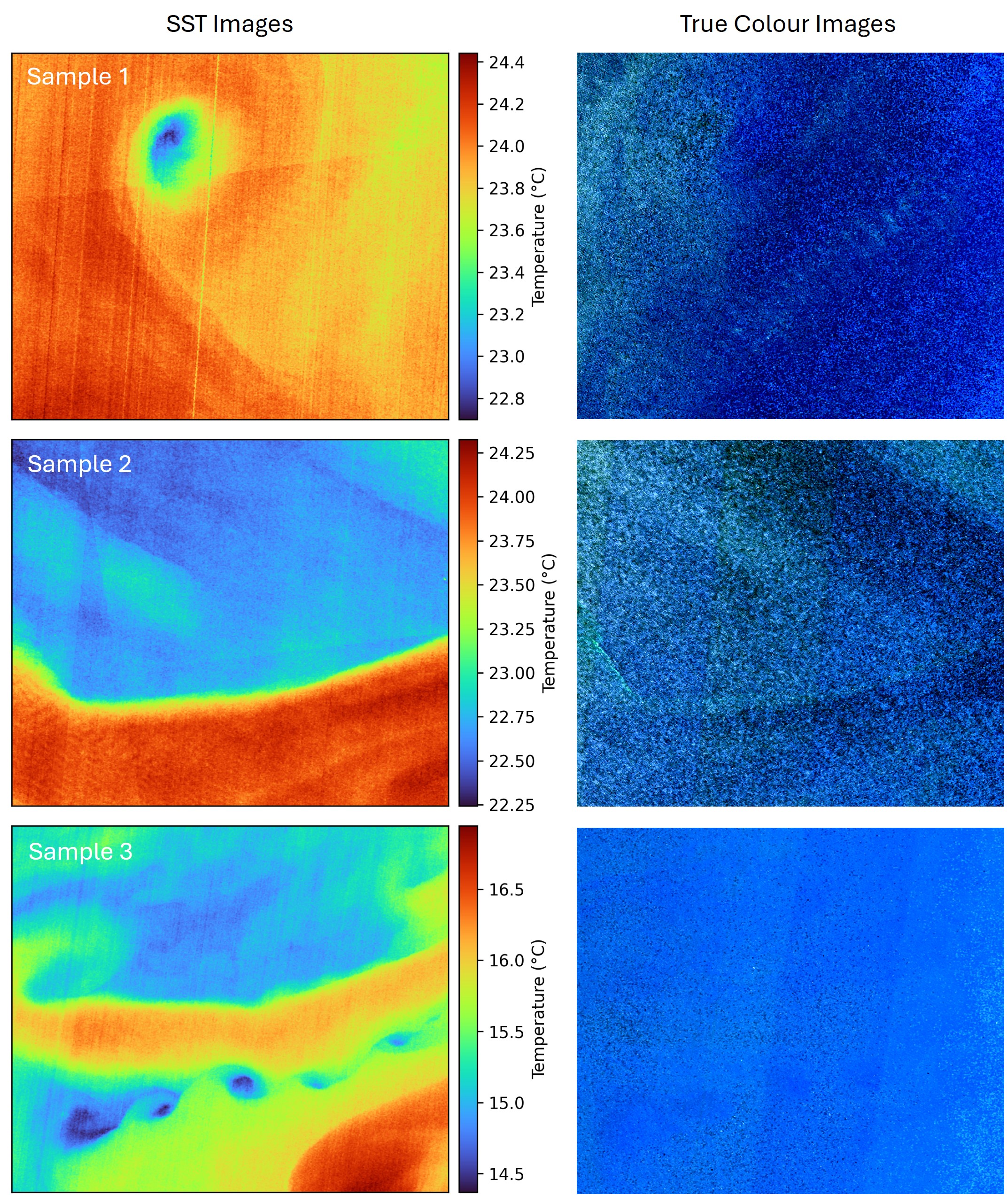}
\caption{Three sample images of sea surface temperature (SST) generated with the proposed workflow (left column), and their spatiotemporally coincident true-colour images (right column).\label{fig:samples}}
\end{figure}

Figure~\ref{fig:samples} presents three randomly sampled SST images generated with the proposed workflow, alongside their aligning (\emph{i.e.}, spatiotemporally coincident) true-colour images. The SST images in the left column revealed the thermal dynamics of sea surface waters, where cooler regions appear in blue and warmer areas in red. In Sample 1, an eddy was identified, likely indicating an upwelling event where colder, deeper waters rise to the surface. A thermal pattern was seen in Sample 2, with a sharp front in SST separating cooler and warmer water masses. From Sample 3, multiple smaller eddies were observed along a temperature gradient, indicating intricate mixing dynamics between warmer and cooler water masses. These thermal patterns reflect the underlying ocean circulation systems that distribute heat, nutrients, and energy in the marine system. While the SST images clearly depict these thermal features, they are less discernible in the aligning true-colour images shown in the right column of Figure~\ref{fig:samples}. However, considering that optical images are valuable in estimating the concentrations of optically active water quality constituents, such as total suspended solids, coloured dissolved organic matter, and chlorophyll \cite{unnithan5123379mapping, peterson2020deep}, it would be worthwhile to synergise SST information with optical data to provide a comprehensive monitoring of marine systems.

\begin{figure}[!htb]
\centering
\includegraphics[width=14cm]{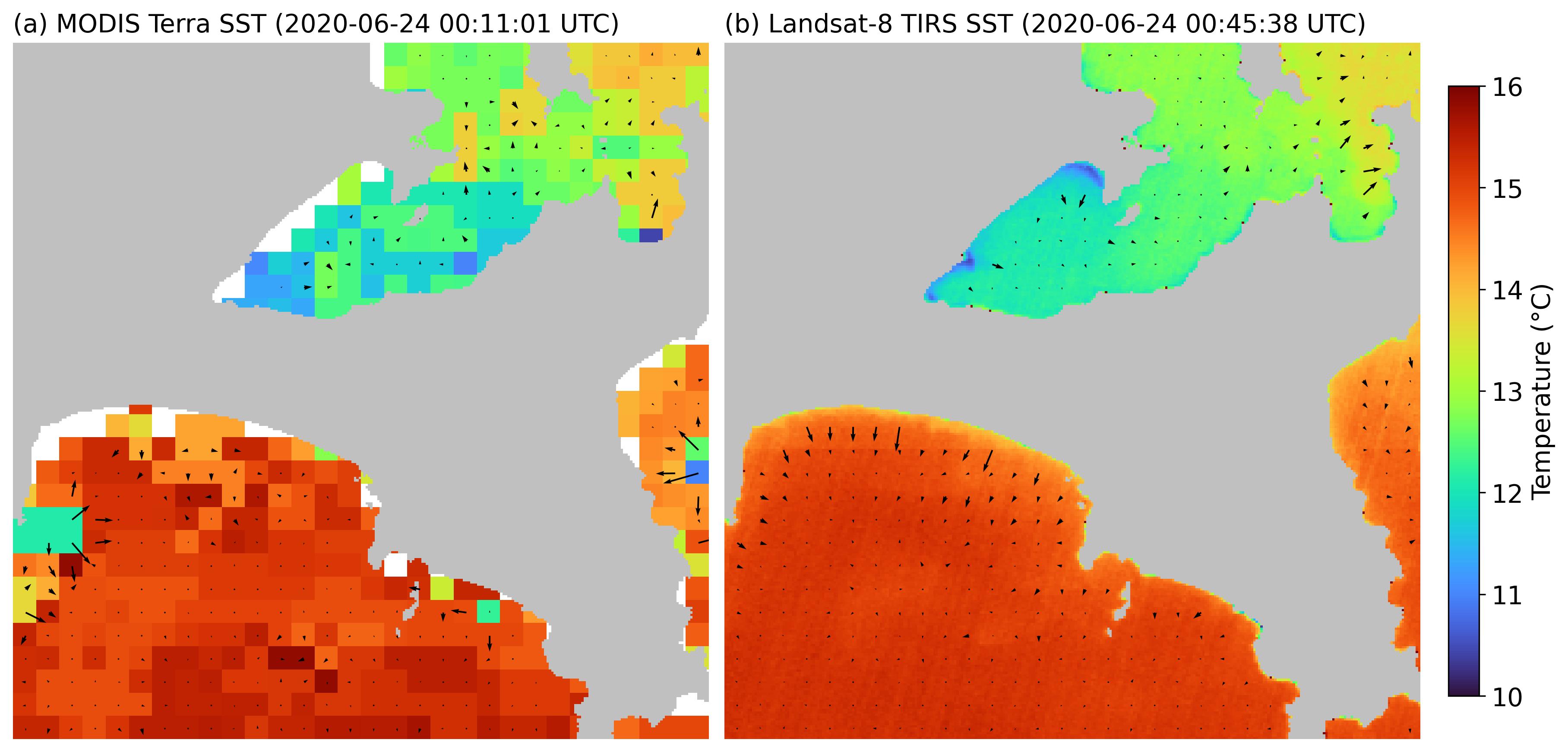}
\caption{Comparison between the sea surface temperature (SST) images (a) recorded by the MODIS Terra sensor (1 km resolution) and (b) derived in this study from Landsat-8 TIRS observations (100 m resolution). The arrows indicate the direction and magnitude of SST gradients.\label{fig:resolution}}
\end{figure}

Figure~\ref{fig:resolution} compares the SST images recorded by the MODIS Terra sensor (Figure~\ref{fig:resolution}a) and derived in this study from the Landsat-8 TIRS observation (Figure~\ref{fig:resolution}b). Both images were captured on June 24, 2020, with a time difference of $\sim$35 min, over waters surrounding the Jussieu Peninsula, South Australia. It was observed from the figure that both images shared a similar spatial pattern of SST, with a comparable range of SST values approximately between 10°C and 16°C. The MODIS Terra SST map (Figure~\ref{fig:resolution}a) displayed a relatively coarse spatial resolution (1 km), with pixelated regions showing distinct blocks of temperature values and less coherent temperature gradients. In contrast, the SST map derived from Landsat-8 TIRS (Figure~\ref{fig:resolution}b) showed a finer spatial resolution (100 m) than the MODIS data (Figure~\ref{fig:resolution}a), displaying spatially more coherent temperature gradients.

\subsection{Spatiotemporal SST Patterns}

\begin{figure}[!htb]
\centering
\includegraphics[width=10cm]{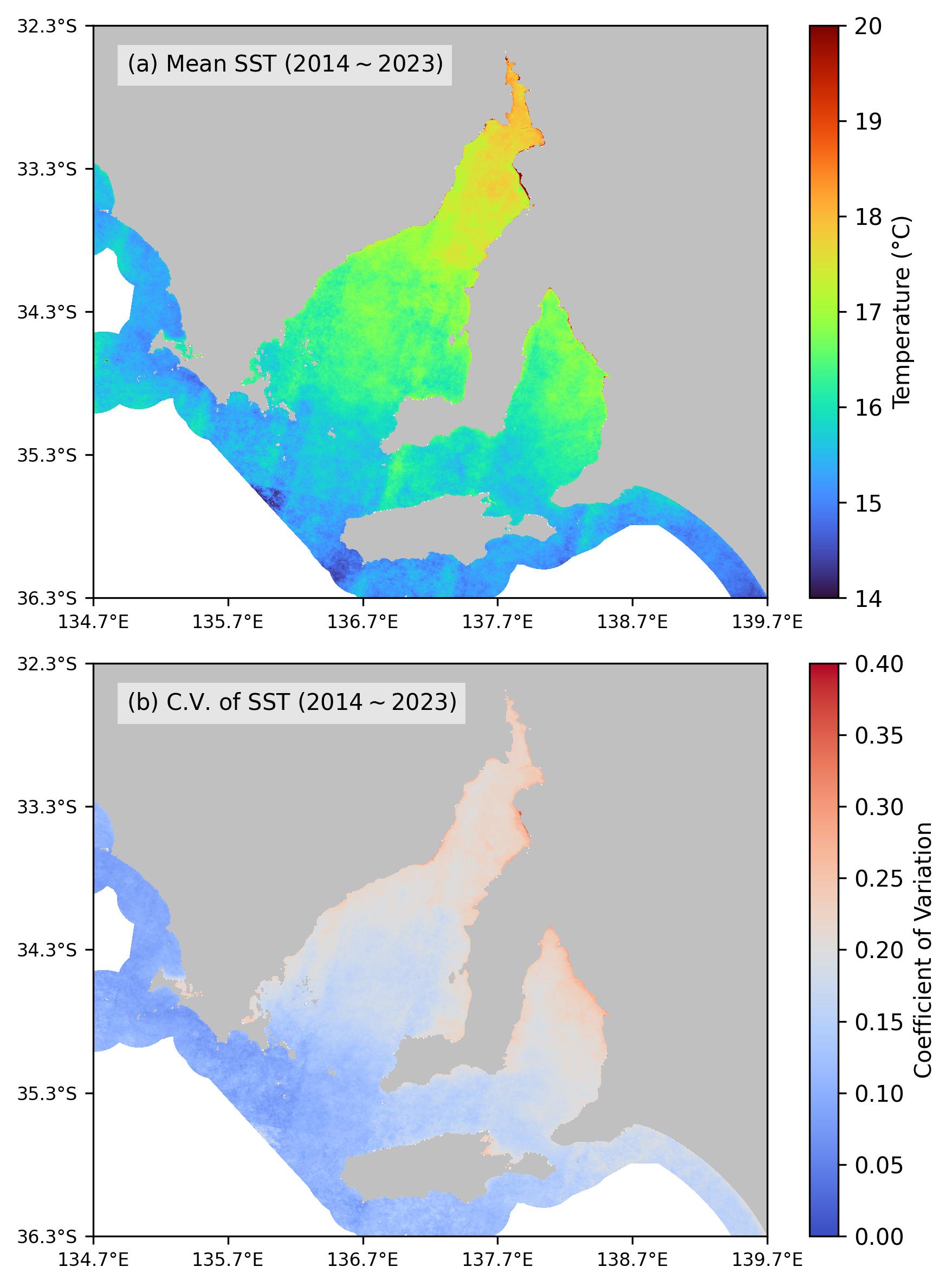}
\caption{(a) The spatial distribution of mean sea surface temperature (SST) averaged over 2014--2023, and (b) the coefficient of variation (C.V.) of SST over the same time period.\label{fig:mean_cv}}
\end{figure}

The spatial distribution of mean SST and the coefficient of variation (C.V.) of SST from 2014 to 2023 are shown in Figure~\ref{fig:mean_cv}. For the mean SST map (Figure~\ref{fig:mean_cv}a), it was seen that the mean temperature in the study area ranges from $\sim$14°C to $\sim$20°C. Warmer waters were observed along the northern coastline, in particular for the Upper Spencer Gulf and Upper St Vincent Gulf regions, while cooler waters were seen in southern areas closer to the open ocean. The temporal variability of SST is illustrated in the C.V. map of SST (Figure~\ref{fig:mean_cv}b). Areas with high C.V. were observed in the Upper Spencer Gulf and Upper St Vincent Gulf regions, suggesting significant temporal variability in these areas. In contrast, the southern areas exhibited lower C.V. values, indicating more stable temperature conditions over time. The spatial distribution of SST and its temporal variability patterns, as given in Figure~\ref{fig:mean_cv}, could inform the local fishery and aquaculture industries for better practices, as the SST plays a key role in influencing the distribution, reproduction, and growth rates of marine species.

\begin{figure}[!htb]
\centering
\includegraphics[width=14cm]{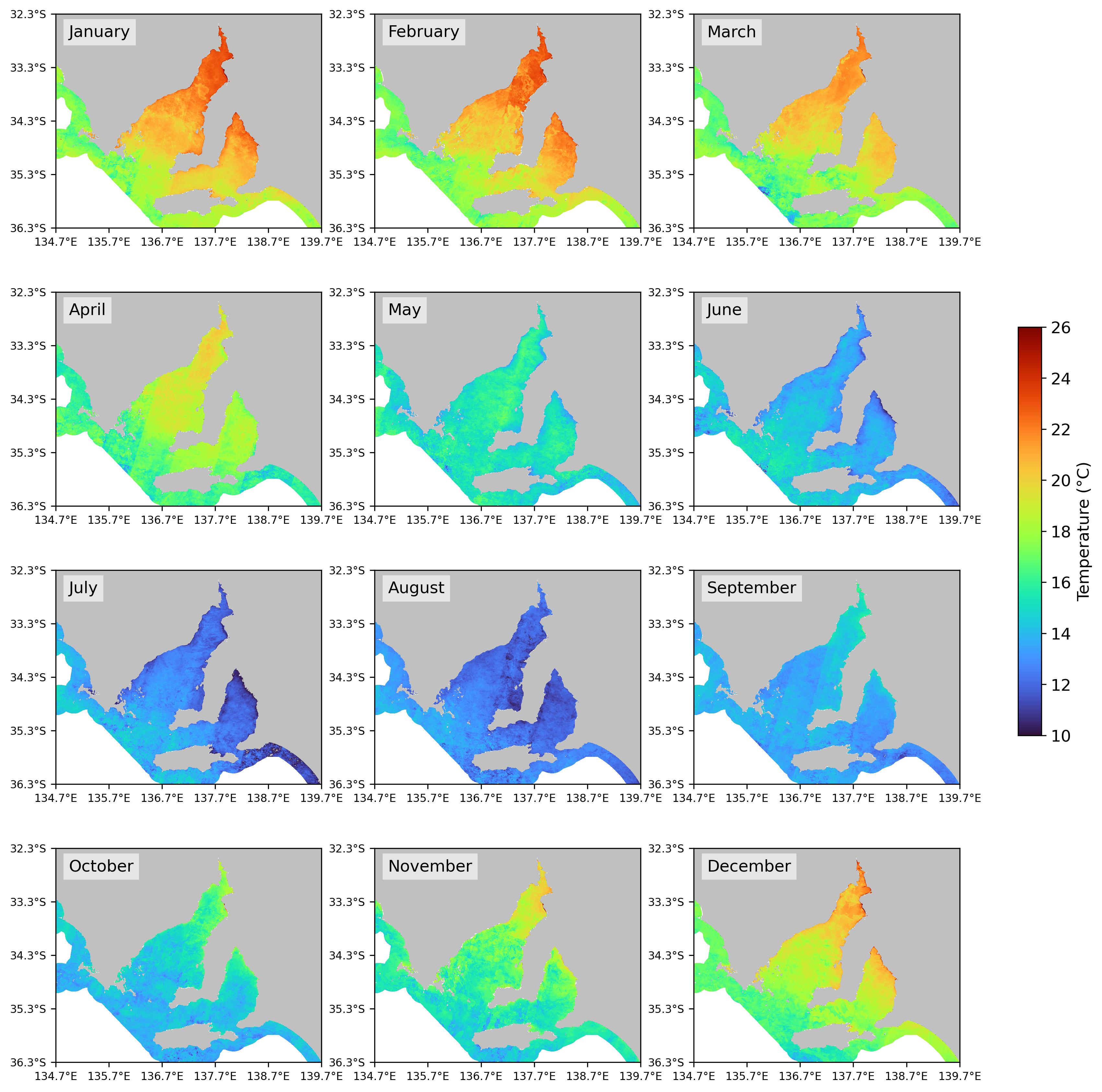}
\caption{Monthly averages of sea surface temperature (SST) in the study area. \label{fig:months}}
\end{figure}

Figure~\ref{fig:months} presents monthly SST maps in the study area, showing the seasonal variability of SST in the study region. A clear cycle of annual warming/cooling was observed, with the warmest temperatures occurring in the summer months (December to March) and the coolest during the winter months. In the summer period, the Upper Spencer Gulf and Upper St Vincent Gulf were observed to have the highest SST values, reaching over 26°C. This warming was seen to be most prominent in the Upper Spencer Gulf and Upper St Vincent Gulf regions close to the coastline. Cooler values were observed during the winter months, especially in July and August, where large portions of the study area exhibit temperatures below 14°C. In particular, the Upper Spencer Gulf and Upper St Vincent Gulf showed cooler temperatures than southern areas closer to the open ocean. The observation from Figure~\ref{fig:months} that the Upper Spencer Gulf and Upper St Vincent Gulf regions showed higher temperatures during summer and lower temperatures during winter explains their higher C.V. values than the southern regions, as depicted in Figure~\ref{fig:mean_cv}b. Given that previous studies suggested that restricted water exchange and shallow bathymetry in semi-enclosed bays could lead to more rapid change in SST (\emph{e.g.}, \cite{jokiel2004global, sanchez2022rapid, lee2005interannual}), the higher summer SST and lower winter SST in Upper Spencer Gulf and Upper St Vincent Gulf could be attributed to those factors. By comparing the monthly mean SST maps, Figure~\ref{fig:months} demonstrates the distinct seasonal SST cycle and its spatial variation. These seasonal patterns are essential for understanding marine ecosystem dynamics and managing fisheries and aquaculture industries that rely on these temperature-dependent environments.

\subsection{Daily Baseline Climatology}

\begin{figure}[!htb]
\centering
\includegraphics[width=14cm]{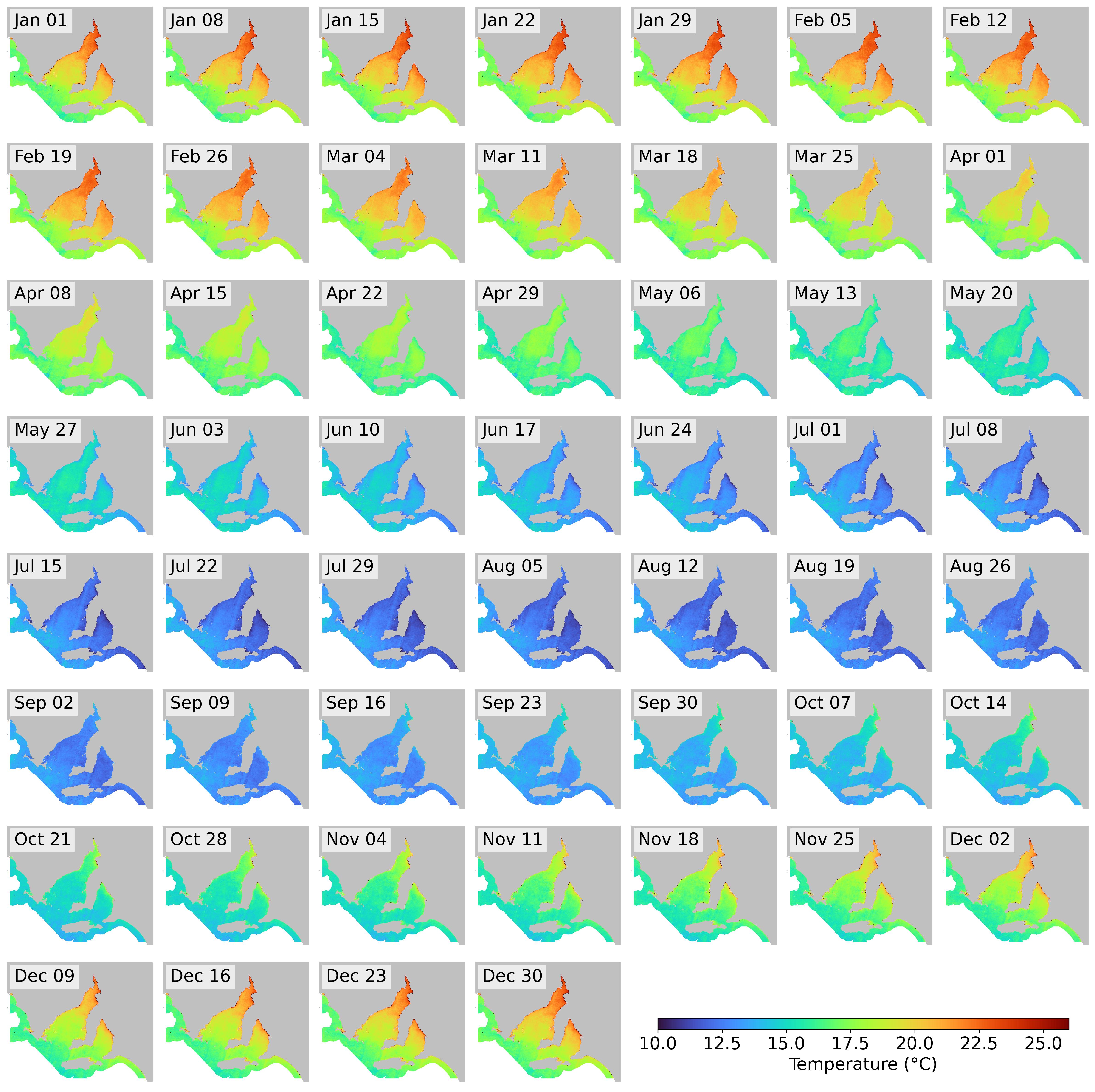}
\caption{Annual baseline climatology of sea surface temperature (SST). Due to space constraints, only weekly interval climatology images are shown here. For access to daily interval climatology images, please refer to the \protect\hyperlink{Data Availability}{Data Availability} section.\label{fig:climatology}}
\end{figure}

The baseline climatology of SST in the study area is shown in Figure~\ref{fig:climatology}. It is worth noting that, due to space constraints, only weekly interval climatology images are shown in the figure, while daily interval climatology images are available via the link provided in the \protect\hyperlink{Data Availability}{Data Availability} section. It was observed from Figure~\ref{fig:climatology} that the climatology follows an annual warming and cooling cycle. Consistent with findings from the monthly mean SST maps in Figure~\ref{fig:months}, the semi-enclosed coastal waters of the gulfs display more pronounced seasonal swings in the SST baseline climatology in Figure~\ref{fig:climatology}, as compared with the open southern regions.   

The SST climatology data derived in this study incorporates the baseline spatiotemporal distribution of SST patterns with a spatial resolution of 100 m and a temporal resolution of 1 day. In terms of spatial resolution, it is higher than commonly used SST climatology data sets, such as the SSTAARS climatology data with 2 km resolution \cite{wijffels2018fine}. Therefore, it allows examining the SST climatology in finer spatial details, especially for near-shore coastal waters where the high spatial resolution near the coastline may reduce the possibility of mixed pixels and reveal small-scale SST features induced by land-water interactions.

It is important to note that the Landsat-8 satellite passes over the study area at approximately 10:00 am local time, meaning that the daily peak/valley temperatures would be missed by the observations due to diurnal warming/cooling. Recalibration of the present climatology data would be necessary to obtain accurate baseline SST values for other times of the day.

\subsection{SST Anomalies}

\begin{figure}[!htb]
\centering
\includegraphics[width=10cm]{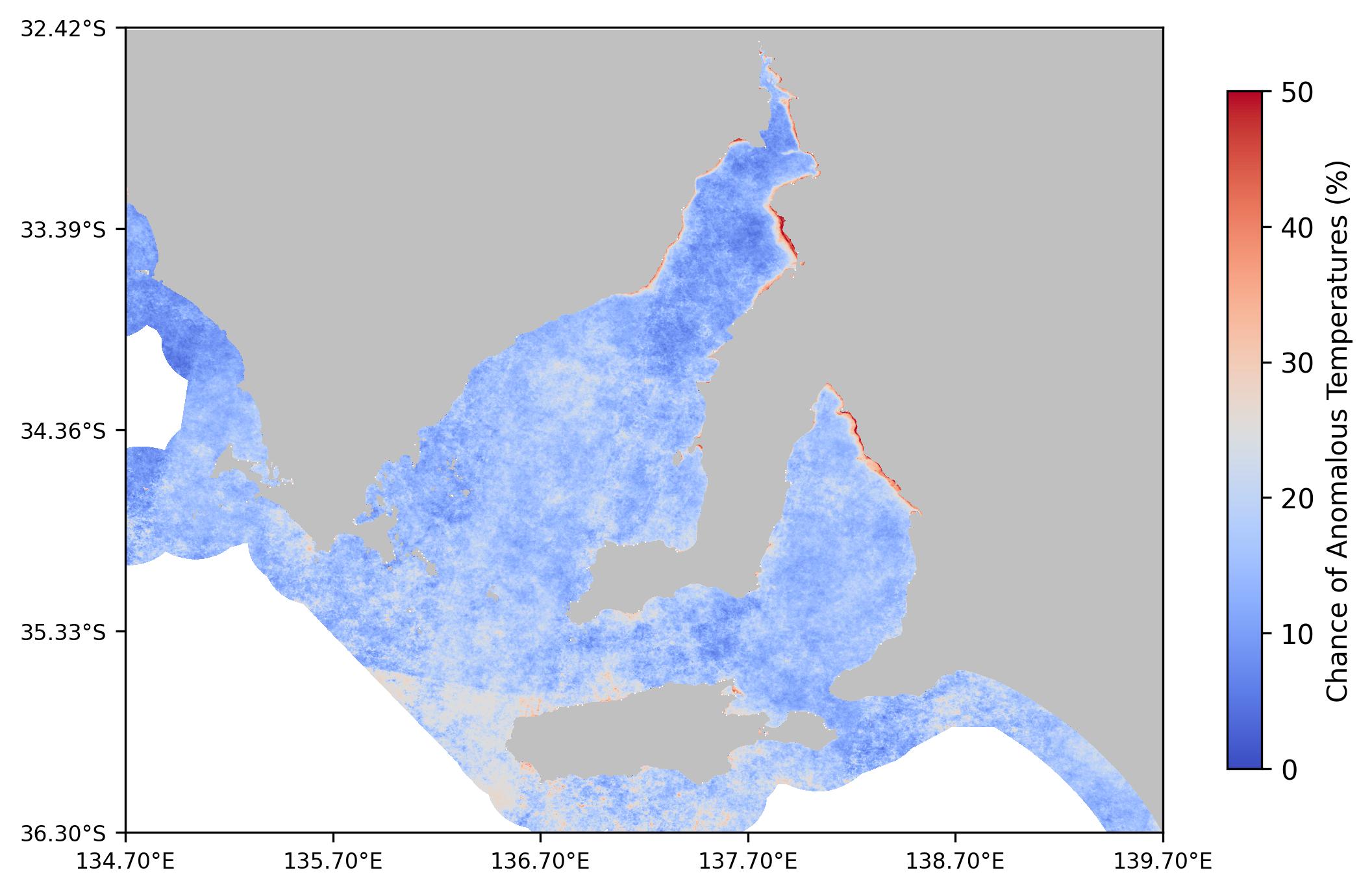}
\caption{The probability of anomalous sea surface temperature (SST) observations in the study area. An anomaly is defined as an observed SST that is 2\textdegree C higher or lower than the baseline climatology. \label{fig:anomaly}}
\end{figure}

Figure~\ref{fig:anomaly} shows the probability of anomalous temperatures in the study area, calculated as the percentage of SST observations from 2014 to 2023 that are greater than or less than 2°C from the baseline climatology. The highest probabilities of anomalous temperatures were observed in the near-shore waters of Upper Spencer Gulf and Upper St Vincent Gulf, with the probability higher than 40\% (Figure~\ref{fig:anomaly}). The high rate of SST anomalies in these areas may be explained by the shallow depth of water (see the bathymetry map of the study area in Figure~\ref{fig:location}). Shallow waters experience more rapid and extreme temperature fluctuations than deeper waters because their lower thermal inertia allows them to quickly heat up and cool down in response to sudden changes in solar radiation and atmospheric temperature, as documented in studies of the Gulf of California \cite{sanchez2022rapid} and Long Island Sound, New York \cite{lee2005interannual}. The restricted water circulation in Upper Spencer Gulf and Upper St Vincent Gulf may also play a role, as limited mixing with open ocean waters can lead to localised warming or cooling---a phenomenon also observed in a study of Hawaii coastal waters \cite{jokiel2004global}. These combined factors make the near-shore, shallow areas in Upper Spencer Gulf and Upper St Vincent Gulf particularly prone to anomalous SST conditions.  

Figure~\ref{fig:anomaly} also reveals medium probabilities of SST anomalies in several isolated areas within the study region, particularly around Kangaroo Island. This pattern aligns with the high cloud coverage observed over these waters, as evidenced in Figure~\ref{fig:count}b. Given that cloud cover is identified as a factor contributing to variations in sea temperature \cite{mclean2014late}, the high frequency of cloud coverage may play a role in the increased likelihood of anomalous SST events  in these waters.

\begin{figure}[!htb]
\centering
\includegraphics[width=14cm]{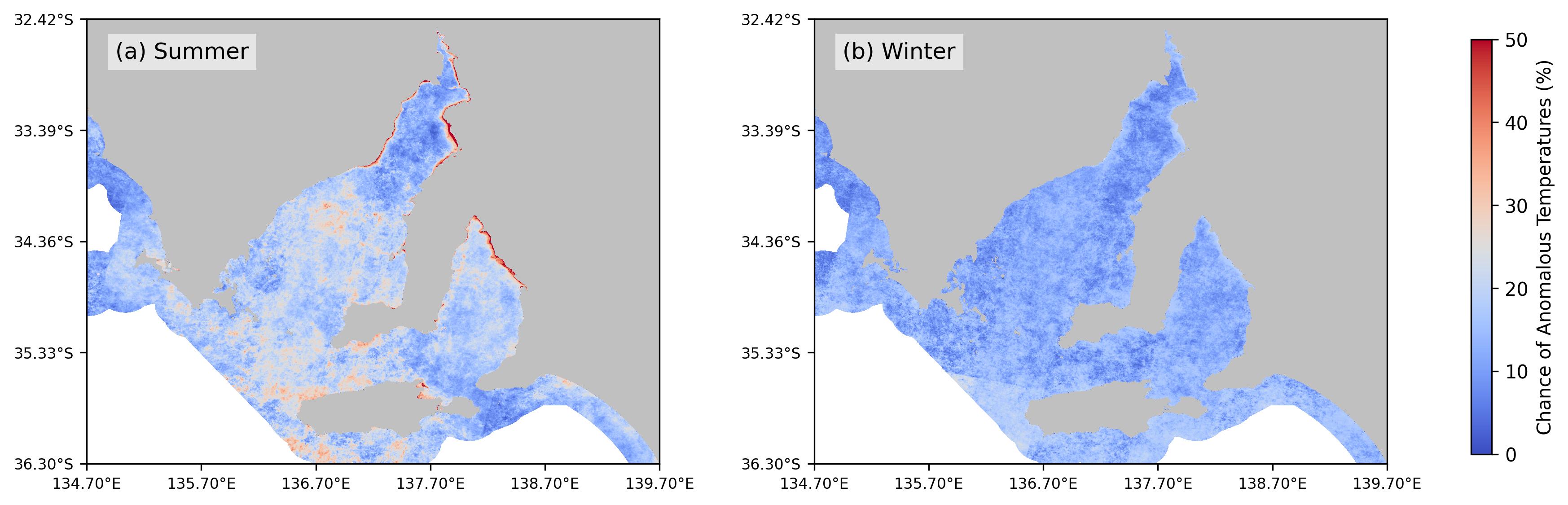}
\caption{The probability of anomalous sea surface temperature (SST) observations in the study area during (a) summer months (January--March and October--December), and (b) winter months (April--September). An anomaly is defined as an observed SST that is 2\textdegree C higher or lower than the baseline climatology. \label{fig:anomaly_summer_winter}}
\end{figure}

To analyse seasonal patterns of SST anomalies, we categorised the anomalous events by the warm months (January--March and October--December) and cool months (April--September), as shown in Figure~\ref{fig:anomaly_summer_winter}. The analysis revealed that most anomaly events occurred during the warm season (Figure~\ref{fig:anomaly_summer_winter}a), such as those detected in the the near-shore areas in Upper Spencer Gulf and Upper St Vincent Gulf, and those surrounding the Kangaroo Island. In comparison, the cool season experienced a lower rate of SST anomalies (Figure~\ref{fig:anomaly_summer_winter}b). These findings suggest that SST during the warm season is more likely to deviate from the baseline climatology than the cool season, indicating more intense/frequent temperature fluctuations during warmer months like summer when the water column is more stratified.

\subsection{Limitations and Prospects}

In this study, we presented an operational method for retrieving SST from the TIRS sensor. We then introduced an algorithm for constructing daily SST climatology, which served as a baseline for identifying anomalous SST events. These methods were applied to temperate coastal waters in South Australia over a ten-year period from 2014 to 2023. While the results demonstrated the potential of Landsat-8 TIRS in high-resolution SST monitoring, there are some limitations and several opportunities for future improvements.

Firstly, this study utilised the Landsat-8 TIRS sensor due to its relatively high spatial resolution (100 m) and over ten years of historical data. This enabled a detailed, long-term analysis of SST in South Australia’s coastal waters. Despite its advantages, Landsat-8 TIRS has a relatively large revisit time of 16 days, and cloud cover may potentially extend data gaps even further. Integrating SST information derived from other satellite sensors, such as Sentinel-3 SLSTR \cite{luo2020validation}, MODIS \cite{xiong2015terra}, and VIIRS \cite{cao2013early}, could help mitigate data gaps and enhance the temporal resolution, although the lower spatial resolutions of these sensor (750 m--1 km) may present an additional challenge for identifying small-scale SST features. The more recently launched TIRS-2 instrument \cite{montanaro2022landsat} onboard Landsat-9 could be leveraged as an additional source of SST data. It is worth noting that integrating SST data from multiple satellite sensors presents several challenges, including spatial and temporal resolution mismatches, varying overpass times, and differences in sensor characteristics and atmospheric correction methods. Addressing these challenges would require approaches such as spatiotemporal harmonisation, cross-sensor calibration, and data fusion techniques.

Secondly, this work focused on monitoring the thermal state of coastal waters. In order to provide an enhanced monitoring of South Australia's coastal environment, it is worth exploring in future studies the synergisation of the thermal observations with information recorded by optical sensors, such as the multispectral instruments Landsat-8 OLI \cite{unnithan5123379mapping, 10854505}, Sentinel-2 MSI \cite{pahlevan2020seamless, guo2018effective} and Sentinel-3 OLCI \cite{pahlevan2020seamless}, and the hyperspectral instruments PACE \cite{gorman2019nasa}, Tanager \cite{rice2024tanager}, EnMAP \cite{chabrillat2024enmap}, and DESIS \cite{krutz2019instrument, guo2023plant}. In addition, sea surface salinity, which can be measured using spaceborne L-band radiometers \cite{kim2023remote}, can also be integrated to enhance insights into the mixing dynamics of fresh and seawater. For example, the remotely sensed salinity data were synergised with bio-optical observations to support water quality monitoring in Florida Bay, USA, as reported in \cite{d2000monitoring}.

Thirdly, this study focused on detecting historical SST anomaly events over the ten-year period from 2014--2023. These detected SST events provide important information on the spatiotemporal distribution of anomalies historically. Implementing real-time anomaly detection techniques, potentially using cloud-based processing platforms, could enable early warnings of significant SST deviations. Such advancements would be beneficial for fisheries, aquaculture, and environmental management by allowing timely responses to anomalous temperature fluctuations that may impact marine ecosystems. However, challenges remain regarding computational resource demands and data latency, which need to be addressed to ensure the reliability and timeliness of real-time anomaly detection systems.

Finally, while this study focused on South Australia’s coastal waters, the proposed approach could be extended to broader geographic regions to support marine and climate research at larger scales. Our valuation results shown in Subsection \ref{ssec:validation_of_sst_retrievals} suggested that the proposed SST retrieval algorithm has the potential to be adapted to regions outside the study area. Applying this methodology to different coastal environments, including tropical and polar regions, would help with a more comprehensive validation of the proposed algorithm.
\section{Conclusion}
\label{sec:conclusion}

This study demonstrated the capability of Landsat-8 TIRS in monitoring SST patterns, climatology, and anomalies in temperate coastal waters off the coast of South Australia over a ten-year period from 2014 to 2023. By developing an operational SST retrieval approach and establishing a high-resolution daily SST climatology, we examined the spatiotemporal patterns of SST and identified regions prone to anomalous temperature events. The high-resolution SST climatology derived in this study represents an improvement over existing datasets, offering enhanced spatial detail for detecting fine-scale temperature variations in coastal waters.

Our results suggested that (1) satellite-derived SST data obtained through the proposed method closely matched in-situ measurements, with an RMSE of approximately 0.38\textdegree C; (2) the Upper Spencer Gulf and Upper St Vincent Gulf, being semi-enclosed regions, experienced more pronounced seasonal temperature variations, with elevated temperatures in summer and lower temperatures in winter relative to open ocean waters; (3) near-shore shallow waters and areas around Kangaroo Island showed a greater likelihood of SST anomalies than other parts of the study area; and (4) anomalous SST events were more prevalent in the warmer months (January--March and October--December) than in the cooler months (April--September). We also showed that the proposed SST retrieval algorithm has the potential to be extended to temperate coastal waters beyond the study area.

These findings provide valuable insights into localised SST dynamics and their implications for fisheries, aquaculture, and marine ecosystem management in temperate coastal waters in South Australia. Future work could focus on integrating information derived from other satellite sensors, incorporating real-time anomaly detection, and extending the approach to broader geographic regions.
\section*{Data Availability}
\label{sec:data_availability}

\begin{itemize}
  \item The SST product generated in this study is accessible via the CSIRO AquaWatch Data Explorer at: \href{https://explorer.adias.aquawatchaus.space/products/sst}{https://explorer.adias.aquawatchaus.space/products/sst}; 
  \item The colour-rendered SST image data can be viewed interactively via the CSIRO AquaWatch Map Portal at:\\ \href{https://map.adias.aquawatchaus.space/#share=s-eBlRT8TGHr8DsQjB}{https://map.adias.aquawatchaus.space/\#share=s-eBlRT8TGHr8DsQjB};
  \item The baseline SST climatology data at daily intervals can be viewed online at: \href{https://drive.google.com/file/d/103RAnCJBONSbCPdafe9XVIqm4o6qdOSw}{https://drive.google.com/file/d/103RAnCJBONSbCPdafe9XVIqm4o6qdOSw}.
\end{itemize}

\section*{Acknowledgements}

The authors would like to thank the anonymous reviewers
for their important and insightful comments for improving this manuscript. The authors also thank Ms Gemma Kerrisk at CSIRO Environment for her assistance in in-situ data acquisition and preparation, and Dr Robert Woodcock and Mr Tisham Dhar at CSIRO Space and Astronomy for their support in data processing and indexing on the EASI ADS platform. The authors acknowledge the New South Wales Government Office of Environment and Heritage (OEH) for providing in-situ sea surface temperature data from the Manly Hydraulics Laboratory Waverider buoys.

This work is supported by Commonwealth Scientific and Industrial Research Organisation (CSIRO) AquaWatch Australia Mission, CSIRO AI4Missions, South Australian Research and Development Institute (SARDI), SmartSat CRC, CSIRO Data61, CSIRO Environment, CSIRO Space and Astronomy, CSIRO Earth Analytics Science and Innovation (EASI) platform, and CSIRO AquaWatch Data Service (ADS).

\textbf{Declaration of using generative artificial intelligence (AI) in the writing process.} During the preparation of this manuscript, the authors utilised GPT-4o to refine language and enhance readability. All AI-assisted content was carefully reviewed and edited as necessary, and the authors take full responsibility for the final version of the publication.

A preprint version of this work is available at \href{https://arxiv.org/abs/2503.05843}{https://arxiv.org/abs/2503.05843}.
\section*{Author Contributions}

Yiqing Guo: Data curation, Formal analysis, Methodology, Visualization, Writing--original draft; Nagur Cherukuru: Conceptualization, Funding acquisition, Supervision, Writing--review \& editing; Eric Lehmann: Conceptualization, Supervision, Writing--review \& editing; Xiubin Qi: Data curation, Writing--review \& editing; Mark Doubell: Funding acquisition, Writing--review \& editing; S. L. Kesav Unnithan: Writing--review \& editing; Ming Feng: Writing--review \& editing.
\section*{Disclosure of interest}

The authors declare no conflict of interest.
\appendix
\section{Atmospheric simulations with libRadtran} 
\label{sec:appendix_a}

Following the approach in \cite{vanhellemont2020automated}, three atmospheric simulations were conducted with the radiative transfer software package \emph{libRadtran} \cite{emde2016libradtran}, to derive the atmospheric upwelling and downwelling radiances, $L_u(\lambda)$ and $L_d(\lambda)$, and the transmittance of atmosphere, $\tau(\lambda)$. It is worth noting that, these simulations were conducted under assumed atmospheric parameters. These assumed parameter values were empirically chosen for the purpose of solving the radiative transfer equation (Eq. (\ref{eq:radiative})), rather than representing the actual atmospheric conditions.

In the first simulation, we assumed the spectral emissivity of water $\varepsilon(\lambda)$ to be 1 for all wavelengths, and the black-body radiance of water $L_s(\lambda)$ to be the spectral radiance of a 290 K black body. We noted the assumed $\varepsilon(\lambda)$ and $L_s(\lambda)$ values as $\varepsilon^{\text{(1)}}(\lambda)$ and $L_s^{\text{(1)}}(\lambda)$, respectively. Under such assumptions, Eq.~(\ref{eq:radiative}) became:
\begin{equation}
\label{eq:sim1}
\begin{aligned}
    L_t^{\text{(1)}}(\lambda) &= \tau(\lambda) \cdot \varepsilon^{\text{(1)}}(\lambda) \cdot L_s^{\text{(1)}}(\lambda) + L_u(\lambda) + \tau(\lambda) \cdot \left(1 - \varepsilon^{\text{(1)}}(\lambda)\right) \cdot L_d(\lambda) \\
    &= \tau(\lambda) \cdot L_s^{\text{(1)}}(\lambda) + L_u(\lambda),
\end{aligned}
\end{equation}
where $L_t^{\text{(1)}}(\lambda)$ is the \emph{libRadtran}-simulated at-sensor radiance under the $\varepsilon^{\text{(1)}}(\lambda)$ and $L_s^{\text{(1)}}(\lambda)$ assumptions. 

In the second simulation, we assumed $\varepsilon(\lambda)$ to be 1 for all wavelengths and $L_s(\lambda)$ to be the spectral radiance of a 300 K black body, noted as $\varepsilon^{\text{(2)}}(\lambda)$ and $L_s^{\text{(2)}}(\lambda)$, respectively, and Eq.~(\ref{eq:radiative}) became:
\begin{equation}
\label{eq:sim2}
\begin{aligned}
    L_t^{\text{(2)}}(\lambda) &= \tau(\lambda) \cdot \varepsilon^{\text{(2)}}(\lambda) \cdot L_s^{\text{(2)}}(\lambda) + L_u(\lambda) + \tau(\lambda) \cdot \left(1 - \varepsilon^{\text{(2)}}(\lambda)\right) \cdot L_d(\lambda) \\
    &= \tau(\lambda) \cdot L_s^{\text{(2)}}(\lambda) + L_u(\lambda),
\end{aligned}
\end{equation}
where $L_t^{\text{(2)}}(\lambda)$ is the corresponding \emph{libRadtran}-simulated at-sensor radiance under the $\varepsilon^{\text{(2)}}(\lambda)$ and $L_s^{\text{(2)}}(\lambda)$ assumptions. By combining Eqs~(\ref{eq:sim1}) and (\ref{eq:sim2}), the transmittance of atmosphere $\tau(\lambda)$ and atmospheric upwelling radiance $L_u(\lambda)$ were derived as:
\begin{equation}
\label{eq:tau}
\tau(\lambda)=\dfrac{L_t^{\text{(2)}}(\lambda)-L_t^{\text{(1)}}(\lambda)}{L_s^{\text{(2)}}(\lambda)-L_s^{\text{(1)}}(\lambda)},
\end{equation}
\begin{equation}
\label{eq:lu}
L_u(\lambda)=L_t^{\text{(1)}}(\lambda)-\dfrac{L_t^{\text{(2)}}(\lambda)-L_t^{\text{(1)}}(\lambda)}{L_s^{\text{(2)}}(\lambda)-L_s^{\text{(1)}}(\lambda)}\cdot L_s^{\text{(1)}}(\lambda).
\end{equation}

In the third simulation, we assumed $\varepsilon(\lambda)$ to be 0.95 and $L_s(\lambda)$ to be 0 (\emph{i.e.}, equivalent to black-body radiance at 0 K), noted as $\varepsilon^{\text{(3)}}(\lambda)$ and $L_s^{\text{(3)}}(\lambda)$, respectively. Then Eq.~(\ref{eq:radiative}) became:
\begin{equation}
\label{eq:sim3}
\begin{aligned}
    L_t^{\text{(3)}}(\lambda) &= \tau(\lambda)\cdot\varepsilon^{\text{(3)}}(\lambda)\cdot L_s^{\text{(3)}}(\lambda) + L_u(\lambda) + \tau(\lambda)\cdot\left[1-\varepsilon^{\text{(3)}}(\lambda)\right]\cdot L_d(\lambda) \\& = L_u(\lambda) + \tau(\lambda)\cdot\left[1-\varepsilon^{\text{(3)}}(\lambda)\right]\cdot L_d(\lambda),
\end{aligned}
\end{equation}
where $L_t^{\text{(3)}}(\lambda)$ is the corresponding \emph{libRadtran}-simulated at-sensor radiance under the $\varepsilon^{\text{(3)}}(\lambda)$ and $L_s^{\text{(3)}}(\lambda)$ assumptions. Taking into consideration of Eqs~(\ref{eq:tau}), (\ref{eq:lu}), and (\ref{eq:sim3}), the atmospheric downwelling radiance $L_d(\lambda)$ was derived as:
\begin{equation}
\label{eq:ld}
L_d(\lambda) = \dfrac{L_t^{\text{(3)}}(\lambda)-L_t^{\text{(1)}}(\lambda)+\dfrac{L_t^{\text{(2)}}(\lambda)-L_t^{\text{(1)}}(\lambda)}{L_s^{\text{(2)}}(\lambda)-L_s^{\text{(1)}}(\lambda)}\cdot L_s^{\text{(1)}}(\lambda)}{\left[1-\varepsilon^{\text{(3)}}(\lambda)\right]\cdot \dfrac{L_t^{\text{(2)}}(\lambda)-L_t^{\text{(1)}}(\lambda)}{L_s^{\text{(2)}}(\lambda)-L_s^{\text{(1)}}(\lambda)}}.
\end{equation}

It is worth noting that the three \emph{libRadtran} simulations were conducted for the thermal spectral range from 9,000 nm to 14,050 nm at 1 nm spectral sampling interval. Therefore, the derived $\tau(\lambda)$, $L_u(\lambda)$, and $L_d(\lambda)$, based on Eqs~(\ref{eq:tau}), (\ref{eq:lu}), and (\ref{eq:ld}), respectively, were of the same spectral range (9,000--14,050 nm) and interval (1 nm).

\section{Calculation of Atmospheric Parameters} 
\label{sec:appendix_b}

For each cell that covers a spatial span of 0.75 degree longitude by 0.75 degree latitude and a temporal span of 3 hours, the corresponding mean sea level air pressure (single level value at sea surface, in Pa) was retrieved from the CAMS EAC4 data set. It was then converted from the single level air pressure at sea surface, to the vertical profile of air pressure at 60 discretised levels (from sea surface to top of atmosphere), based on the 60-level vertical discretisation model as defined in \cite{74321} and the converting coefficients for each model level as provided in CAMS EAC4:
\begin{equation}
\label{eq:pressure}
p^{(i)}=a^{(i)}+p_0\cdot b^{(i)},
\end{equation}
where $p_0$ is the air pressure at sea surface; $p^{(i)}$ is the air pressure at the $i$th model level; and $a^{(i)}$ and $b^{(i)}$ are the corresponding converting coefficients for that model level. 

Cell-specific retrievals of atmosphere conditions from the CAMS EAC4 data set also included air temperature $T^{(i)}$ (in unit of K), specific humidity $q^{(i)}$ (in unit of kg/kg), ozone concentration $O_3^{(i)}$ (in unit of kg/kg), and nitrogen dioxide concentration $NO_2^{(i)}$ (in unit of kg/kg). All these parameters were provided as vertical profiles at the 60 model levels from sea surface to top of atmosphere, with $i$ ranging from 1 to 60 in their notations.

The altitude of each model level was calculated from the air pressure at sea surface $p_0$ and the vertical profiles of air pressure $p^{(i)}$ and air temperature $T{(i)}$ via the Barometric formula:
\begin{equation}
\label{eq:altitude}
h^{(i)}=-\dfrac{\text{ln}\left(\dfrac{p^{(i)}}{p_0}\right)\cdot R\cdot T^{(i)}}{g\cdot M}+h_0,
\end{equation}
where $h^{(i)}$ is the altitude at the $i$th model level; $h^{(i)}$=0 m is the altitude of sea surface; $g$=9.8067 m/s\textsuperscript{2} is the gravitational acceleration; $R$=8.31446 J/(K·mol) is the universal gas constant; and $M^{(i)}$ is the molar mass of air at the $i$th model level, which was calculated as the combination of molar mass of dry air and water vapour weighted by the specific humidity of the same model level $q^{(i)}$:
\begin{equation}
\label{eq:molar_mass}
M^{(i)}=M_\text{d}\cdot \left(1-q^{(i)}\right)+M_\text{v}\cdot q^{(i)},
\end{equation}
where $M_\text{d}$=0.0289652 kg/mol and $M_\text{v}$=0.018016 kg/mol are the molar mass of dry air and water vapour, respectively.

The air density $\rho^{(i)}$ at the $i$th model level was calculated based on the Ideal Gas Law by treating the atmosphere as a mixture of dry air and water vapour:
\begin{equation}
\label{eq:density}
\rho^{(i)} = \dfrac{p_\text{d}^{(i)}\cdot M_\text{d} + p_\text{v}^{(i)}\cdot M_\text{v}}{R\cdot T^{(i)}},
\end{equation}
where $p_\text{d}^{(i)}$ and $p_\text{v}^{(i)}$ are the partial pressure of dry air and water vapour, respectively, at the $i$th model level. The water vapour pressure $p_\text{v}^{(i)}$ was calculated as \cite{murray1967computation,shaman2009absolute}:
\begin{equation}
\label{eq:pressure_water_vapour}
p_\text{v}^{(i)}=\dfrac{q^{(i)}\cdot p^{(i)}}{\dfrac{M_\text{v}}{M_\text{d}} + \left(1-\dfrac{M_\text{v}}{M_\text{d}} \right)\cdot q^{(i)}},
\end{equation}
where $p^{(i)}$ is air pressure and $q^{(i)}$ is specific humidity at the $i$th model level, both of which were provided by CAMS EAC4. Given $p_\text{v}^{(i)}$, the dry air pressure $p_\text{d}^{(i)}$ can be calculated as:
\begin{equation}
\label{eq:pressure_dry_air}
p_\text{d}^{(i)}=p^{(i)}-p_\text{v}^{(i)}.
\end{equation}

Nitrogen ($N_2$), oxygen ($O_2$), and carbon dioxide ($CO_2$) constitute 75.523\%, 23.133\%, and 0.053\%, respectively, of dry air by mass ratio (\emph{i.e.}, in unit of kg/kg). Although the compositions of these gasses in dry air don't change with altitude, their compositions in total air do due to the vertical variations of water vapour content. To this end, the nitrogen concentration $N_2^{(i)}$ (in unit of kg/kg), oxygen concentration $O_2^{(i)}$ (in unit of kg/kg), and carbon dioxide concentration $CO_2^{(i)}$ (in unit of kg/kg) at the $i$th model level were calculated with the specific humidity of that model level $q^{(i)}$ being considered:
\begin{align}
\label{eq:nitrogen}
N_2^{(i)} &= \left(1 - q^{(i)}\right) \times 0.75523, \\
\label{eq:oxygen}
O_2^{(i)} &= \left(1 - q^{(i)}\right) \times 0.23133, \\
\label{eq:carbon_dioxide}
CO_2^{(i)} &= \left(1 - q^{(i)}\right) \times 0.00053.
\end{align}

\bibliographystyle{tfq}
\bibliography{ref}

\end{document}